%% file: paper.tex
\documentstyle{mn}
\input{psfig.sty}
\input{macro1}

\begin{document}

%%%%%%%%%%%%%%%%%%%%%%%%%%%%%%%%%%%%%%%%%%%%%%%%%%%%%%%%%%%%%%%%%%%%%%%%%%

\title[Galaxy Occupation Statistics of Dark Matter Haloes: Observational Results]
      {Galaxy Occupation Statistics of Dark Matter Haloes: Observational Results}

\author[Yang, Mo, Jing, van den Bosch]
       {Xiaohu Yang$^{1}$, H.J. Mo$^{1}$, Y.P. Jing$^{2}$, 
        Frank C. van den Bosch$^{3}$ 
        \thanks{E-mail: xhyang@astro.umass.edu}\\
      $^1$Department of Astronomy, University of Massachusetts,
          Amherst MA 01003-9305, USA\\
      $^2$Shanghai Astronomical Observatory; the Partner Group of MPA,
          Nandan Road 80, Shanghai 200030, China\\
      $^3$Department of Physics, Swiss Federal Institute of
          Technology, ETH H\"onggerberg, CH-8093, Zurich,
          Switzerland}

%%%%%%%%%%%%%%%%%%%%%%%%%%%%%%%%%%%%%%%%%%%%%%%%%%%%%%%%%%%%%%%%%%%%%%%%%%

\date{}

%\pagerange{\pageref{firstpage}--\pageref{lastpage}}
%\pubyear{2004}

\maketitle

\label{firstpage}

%%%%%%%%%%%%%%%%%%%%%%%%%%%%%%%%%%%%%%%%%%%%%%%%%%%%%%%%%%%%%%%%%%%%%%%%%%

\begin{abstract}  
  We study the occupation statistics of galaxies in dark matter haloes
  using  galaxy  groups  identified  from the  2-degree  Field  Galaxy
  Redshift  Survey with  the  halo-based group  finder  of Yang  \etal
  (2004b).  The  occupation distribution is  considered separately for
  early and  late type galaxies,  as well as  in terms of  central and
  satellite  galaxies. The  mean  luminosity of  the central  galaxies
  scales  with halo  mass  approximately as  $L_c\propto M^{2/3}$  for
  haloes  with  masses  $M<  10^{13}h^{-1}\msun$, and  as  $L_c\propto
  M^{1/4}$  for  more  massive  haloes.  The  characteristic  mass  of
  $10^{13}  h^{-1} \Msun$  is  consistent with  the  mass scale  where
  galaxy  formation  models suggest  a  transition  from efficient  to
  inefficient cooling. Another characteristic halo mass scale, 
  $M\sim 10^{11}h^{-1}\msun$, which cannot be probed directly by our groups, 
  is inferred from the conditional luminosity function (CLF) that 
  matches the observed galaxy luminosity function and clustering. 
  For a halo  of given mass, the distribution of
  $L_c$  is  rather  narrow.   Detailed comparison  with  mock  galaxy
  redshift  surveys  indicates  this  implies a  fairly  deterministic
  relation  between  $L_c$ and  halo  mass.   The satellite  galaxies,
  however, are  found to follow  a Poissonian number  distribution, in
  excellent agreement  with the  occupation statistics of  dark matter
  subhaloes.  This provides strong  support for the standard lore that
  satellite  galaxies reside  in subhaloes.   The central  galaxies in
  low-mass  haloes  are mostly  late  type  galaxies,  while those  in
  massive  haloes are  almost all  early types.   We also  measure the
  CLF of galaxies in haloes of given
  mass.   Over the mass  range that  can be  reliably probed  with the
  present data ($13.3 \lta {\rm log}[M/(h^{-1}\Msun)] \lta 14.7$), the
  CLF is  reasonably well  fit by a  Schechter function.   Contrary to
  recent claims  based on semi-analytical models  of galaxy formation,
  the presence of  central galaxies does not show up  as a strong peak
  at  the  bright  end  of   the  CLF.  The  CLFs  obtained  from  the
  observational data are in good agreement with the CLF model obtained
  by  matching  the   observed  luminosity  function  and  large-scale
  clustering  properties  of  galaxies  in the  standard  $\Lambda$CDM
  model.
\end{abstract}

%%%%%%%%%%%%%%%%%%%%%%%%%%%%%%%%%%%%%%%%%%%%%%%%%%%%%%%%%%%%%%%%%%%%%%%%%%


%%%%%%%%%%%%%%%%%%%%%%%%%%%%%%%%%%%%%%%%%%%%%%%%%%%%%%%%%%%%%%%%%%%%%%%%%%

\section{Introduction}

According  to the  current paradigm  of structure  formation, galaxies
form and reside  inside extended cold dark matter  (CDM) haloes. These
haloes  are virialized  clumps that  formed through  the gravitational
instability of the  cosmic density field, and have  typical sizes that
are  much smaller  than their  mean  spatial separation.   One of  the
ultimate  challenges   in  astrophysics   is  to  obtain   a  detailed
understanding  of  how  galaxies  with different  physical  properties
occupy  dark  matter haloes  of  different  mass.   This link  between
galaxies and dark  matter haloes is an imprint  of various complicated
physical processes related to  galaxy formation, such as gravitational
instability, gas cooling, star formation, merging, tidal stripping and
heating,   and  a   variety   of  feedback   processes.   A   detailed
quantification of this link is therefore pivotal for our understanding
of galaxy  formation and evolution within the  CDM cosmogony. Although
the statistical  link itself does  not give a physical  explanation of
how  galaxies form and  evolve, it  provides important  constraints on
these processes  and on how their  efficiencies scale with  halo mass. 

To  quantify  the  relationship  between  haloes  and  galaxies  in  a
statistical way, it has become customary to specify the so-called halo
occupation distribution,  $P(N \vert M)$, which  gives the probability
to find  $N$ galaxies  (with some specified  properties) in a  halo of
mass $M$.  This occupation  distribution can be constrained using data
on the  clustering properties of galaxies, as  it completely specifies
the galaxy  bias, and  has been used  extensively to study  the galaxy
distribution in dark matter haloes  and the galaxy clustering on large
scales (Jing, Mo \& B\"orner 1998; Peacock \& Smith 2000; Seljak 2000;
Scoccimarro  \etal  2001; Jing,  B\"orner  \&  Suto  2002; Berlind  \&
Weinberg 2002;  Bullock, Wechsler  \& Somerville 2002;  Scranton 2002;
Kang  \etal   2002;  Marinoni  \&  Hudson  2002;   Zheng  \etal  2002;
Magliocchetti  \&  Porciani 2003;  Berlind  \etal  2003; Zehavi  \etal
2004a,b; Zheng \etal 2004).

Since  individual  galaxies  are  not featureless  objects,  but  have
diverse  intrinsic properties,  a  more useful  halo occupation  model
should contain  some information regarding the  physical properties of
the galaxies.  A significant step  in this direction has been taken by
Yang,  Mo \&  van den  Bosch (2003b)  and van  den Bosch,  Yang  \& Mo
(2003a), who modelled the halo occupation as a function of both galaxy
luminosity and  type (see  also Vale \&  Ostriker 2004 for  a somewhat
different approach).   In particular, they  introduced the conditional
luminosity function (CLF), $\Phi(L \vert M) {\rm d}L$, which gives the
average number  of galaxies  with luminosity $L  \pm {\rm  d}L/2$ that
reside in a halo of mass $M$. As shown by Yang \etal (2003b), once the
galaxy luminosity  function and the galaxy correlation  amplitude as a
function of  luminosity are known, tight constraints  on $\Phi(L \vert
M)$ can  be obtained.  Detailed comparisons with  additional data from
the 2dFGRS, the  Sloan Digital Sky Survey (SDSS)  and DEEP2 have shown
that the resulting halo occupation models can reproduce a large number
of  observations regarding  the  galaxy distribution  at low  redshift
(Yan, Madgwick \&  White 2003; Yang \etal 2004a; Mo  et al. 2004; Wang
\etal 2004;  Zehavi \etal 2004b; Yan,  White \& Coil  2004).  This not
only implies  that these  occupation distributions provide  a reliable
description of the connection between galaxies and CDM haloes, it also
implies that  the standard $\Lambda$CDM model is  a good approximation
to  the  real Universe.   After  all,  the  abundances and  clustering
properties of dark matter haloes are cosmology dependent, and matching
the data with occupation models  is only possible for a restricted set
of  cosmological parameters  (Zheng \etal  2002; van  den  Bosch \etal
2003b; Rozo, Dodelson \& Frieman 2004; Abazajian \etal 2004).

An important shortcoming of  these occupation models, however, is that
the   results  are  not   completely  model   independent.   Typically
assumptions have  to be made  regarding the functional form  of either
$P(N \vert M)$ or $\Phi(L \vert  M)$.  For example, in our work on the
CLF we  have always assumed that  it is well described  by a Schechter
function  (Yang  \etal 2003b;  van  den  Bosch  \etal 2003a,  2004b).  
Recently, however,  the validity of this assumption  was questioned by
Zheng \etal (2004), based on  a study of the conditional baryonic mass
function (similar to the CLF  but with luminosity replaced by baryonic
mass) in semi-analytical models of galaxy formation.  Note that in all
halo  occupation studies  to date,  the occupation  distributions have
been determined in an indirect way: the free parameters of the assumed
functional form  are constrained using  {\it statistical} data  on the
abundance  and  clustering  properties   of  the  galaxy  population.  
Ideally, however, one would determine the occupation distribution more
directly, by using  a method that can determine  which galaxies belong
to the  same dark  matter halo.  If  such a  method can be  found, the
occupation  statistics, including  the CLF,  can be  obtained directly
from the data without the need to make any assumptions.

In this paper we perform such a direct determination of the occupation
statistics using  the halo-based group finder developed  by Yang \etal
(2004b).  Detailed  tests with mock galaxy catalogues  have shown that
this group finder is very successful in associating galaxies according
to  their  common  dark   matter  haloes  (Yang  \etal  2004b,c).   In
particular,  the group  finder  performs reliably  not  only for  rich
systems,  but  also  for  poor  systems,  including  isolated  central
galaxies  in  low-mass  haloes,   making  it  possible  to  study  the
galaxy-halo connection for a wide range of different systems.  In this
paper, we  use the  sample of galaxy  groups obtained from  the 2dFGRS
with this  group finder to  study the galaxy occupation  statistics in
dark matter haloes  as a function of halo  mass, galaxy luminosity and
type,  and in  terms  of  both central  and  satellite galaxies.   The
arrangement of  the paper is as  follows: In Section~\ref{sec:catalog}
we describe the  data and the mock surveys used in  the present paper. 
Sections~\ref{sec:property} and~\ref{sec:CLF}  presents our results on
the halo  occupation distribution and on the  CLF.  Further discussion
and     a    summary     of     our    results     are    given     in
Section~\ref{sec:conclusion}.

\section{Group Catalogues}
\label{sec:catalog}

\subsection{Galaxy Groups in the 2dFGRS}

Here we  briefly describe  the group catalogues  used in  the analyzes
that  follow.   The  construction  of  these catalogues,  as  well  as
numerous  tests regarding  the  performance of  the  group finder,  is
described in Yang \etal (2004b, hereafter YMBJ ) to which we refer the
interested reader for details.
\begin{figure*}
\centerline{\psfig{figure=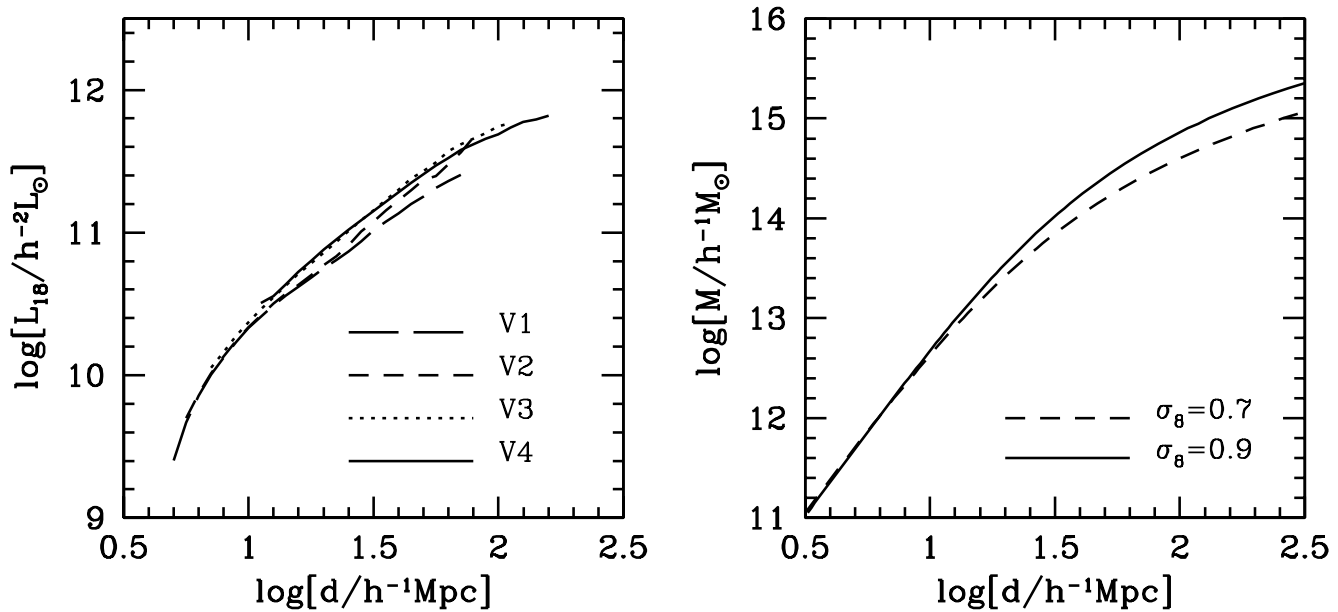,width=\hdsize}} 
\caption{The left-hand  panel shows  the  relation between  the group  
  luminosity $L_{18}$ and the mean separation $d$ of all groups with a
  group luminosity  larger than $L_{18}$.   Different lines correspond
  to different  `mass-limited' group samples obtained  from the 2dFGRS
  (see Table~1).  The small differences at large $d$ are due to cosmic
  variance.  The right-hand panel  shows the relation between the halo
  mass $M$ and mean halo separation $d$ derived from the mass function
  of  dark  matter  haloes   for  two  $\Lambda$CDM  cosmologies  with
  different $\sigma_8$, as indicated. Throughout this paper we compute
  halo masses  from group  luminosities as follows:  for a  group with
  given  $L_{18}$ we  determine the  mean separation  $d$  between all
  groups with  a group  luminosity larger than  $L_{18}$, and  use the
  panel on the  right to determine the halo  mass $M$ that corresponds
  to this $d$ for the cosmology under consideration. }
\label{fig:M_L_d}
\end{figure*}

The basic idea behind the group finder developed by YMBJ is similar to
that of the matched filter algorithm developed by Postman \etal (1996)
(see also Kepner  \etal 1999; White \& Kochanek  2002; Kim \etal 2002;
Kochanek \etal  2003; van den  Bosch \etal 2004a,b), although  it also
makes use of  the galaxy kinematics.  The group  finder starts with an
assumed  mass-to-light  ratio  to  assign  a tentative  mass  to  each
potential  group   (identified  using  the   Friends-of-Friends  (FOF)
method).   This  mass  is  used  to estimate  the  size  and  velocity
dispersion of the underlying halo  that hosts the group, which in turn
is  used to  determine  group membership  (in  redshift space).   This
procedure  is  iterated  until  no  further  changes  occur  in  group
memberships.  The performance of the  group finder was tested in terms
of  completeness of  true  members and  contamination by  interlopers,
using detailed mock galaxy redshift surveys.  The average completeness
of  individual groups is  $\sim 90$  percent and  with only  $\sim 20$
percent  interlopers.  Furthermore, the  resulting group  catalogue is
insensitive  to  the initial  assumption  regarding the  mass-to-light
ratios, and the group finder  is more successful than the conventional
FOF  method   (e.g.,  Eke  \etal  2004  and   references  therein)  in
associating galaxies according to their common dark matter haloes.

In YMBJ  we used this  group finder to  identify galaxy groups  in the
final  public data  release of  the 2dFGRS.   This redshift  sample of
galaxies  contains about  $250,000$  galaxies and  is  complete to  an
extinction-corrected apparent magnitude of $b_J\approx 19.45$ (Colless
\etal 2001).  When identifying  galaxy groups, we restricted ourselves
to  galaxies  with redshifts  $0.01\leq  z  \leq  0.20$ in  the  North
Galactic Pole  (NGP) and the  South Galactic Pole (SGP)  regions. Only
galaxies with redshift quality parameter  $q \geq 3$ and with redshift
completeness $>0.8$  were used. This  left a grand total  of $151,280$
galaxies with a sky coverage of $1124 ~{\rm deg}^2$. From this sample,
YMBJ  obtained a group  catalogue that  contains $78,708$  systems, of
which 7251 are binaries, 2343  are triplets, and 2502 are systems with
four or more members.  In what  follows we use this group catalogue to
determine the halo occupation statistics of the 2dFGRS.

\subsection{Mock Group Catalogues}
\label{sec:mock}

In testing  the halo-based group finder,  YMBJ used a  set of detailed
mock galaxy redshift surveys (hereafter MGRSs). Here we use these same
MGRSs for  comparison with  the 2dFGRS. For  the present  analysis, we
correct these  MGRSs for close  pair incompleteness that  arises from
fiber collisions  and from the  fact that nearby galaxies  overlap (so
that  they are identified  as a  single galaxy,  rather than  a galaxy
pair). The method used to correct  our MGRSs for both these effects is
described in  detail in  van den Bosch  \etal (2004b). Note  that this
close-pair incompleteness has  only a minor impact on  our results: in
other words, if we were not to correct for these effects, it would not
impact any of  our main conclusions. In what follows,  we give a brief
description about  how these MGRSs  are constructed, and we  refer the
reader  to Yang  \etal (2004a)  and van  den Bosch  \etal  (2004a) for
details.

The mock surveys  are constructed by populating dark  matter haloes in
large  numerical simulations with  galaxies of  different luminosities
and  different types.   The simulations  correspond to  a $\Lambda$CDM
concordance  cosmology  with  $\Omega_m=0.3$,  $\Omega_{\Lambda}=0.7$,
$h=H_0/(100  \kmsmpc)=0.7$ and  with a  scale-invariant  initial power
spectrum with normalization $\sigma_8=0.9$, and all MGRSs discussed in
this paper are therefore only  valid for this particular cosmology. To
populate the dark matter haloes  with galaxies we use the CLF. Because
of  the  mass  resolution  of  the  simulations  and  because  of  the
completeness limit of the 2dFGRS  we adopt a minimum galaxy luminosity
of $L_{\rm  min} =  10^{7} h^{-2} \Lsun$  throughout.  The  {\it mean}
number of galaxies with $L \geq L_{\rm min}$ that resides in a halo of
mass $M$ is given by
\begin{equation}
\label{averN}
\langle N \rangle_M = \int_{L_{\rm min}}^{\infty} \Phi(L \vert M)
\, {\rm d}L
\end{equation}
In  order  to Monte-Carlo  sample  occupation  numbers for  individual
haloes one  requires the full probability distribution  $P(N \vert M)$
(with $N$ an integer) of  which $\langle N \rangle_M$ gives the mean,
i.e.,
\begin{equation}
\label{meanNint}
\langle  N \rangle_M = \sum_{N=0}^{\infty} N \, P(N \vert M)
\end{equation}

We use the  results of Kravtsov \etal (2004a), who  has shown that the
number  of {\it subhaloes}  follows a  Poisson distribution.   In what
follows  we   differentiate  between  satellite   galaxies,  which  we
associate  with these  dark  matter subhaloes,  and central  galaxies,
which we  associate with the host  halo. The total  number of galaxies
per halo is  the sum of $N_{\rm cen}$, the  number of central galaxies
which is either one or zero, and $N_{\rm sat}$, the (unlimited) number
of satellite galaxies.  We assume that $N_{\rm sat}$ follows a Poisson
distribution  and  require   that  $N_{\rm  sat}=0$  whenever  $N_{\rm
  cen}=0$.   The halo  occupation  distribution is  thus specified  as
follows: if  $\langle N \rangle_M \leq  1$ then $N_{\rm sat}  = 0$ and
$N_{\rm cen}$  is either  zero (with  probability $P =  1 -  \langle N
\rangle_M$) or one  (with probability $P = \langle  N \rangle_M$).  If
$\langle  N \rangle_M  > 1$  then  $N_{\rm cen}=1$  and $N_{\rm  sat}$
follows the Poisson distribution
\begin{equation}
\label{poisson}
P(N_{\rm sat} \vert M) = {\rm e}^{-\mu} 
{\mu^{N_{\rm sat}} \over N_{\rm sat}!}\,,
\end{equation}
with $\mu = \langle N_{\rm sat}  \rangle_M = \langle N \rangle_M - 1$.

We  follow Yang  \etal (2004a)  and van  den Bosch  \etal  (2004b) and
assume that the  central galaxy is the brightest galaxy  in each halo. 
Its luminosity  is drawn from  $\Phi(L \vert M)$ with  the restriction
that $L > L_1$ with $L_1$ defined by
\begin{equation}
\label{Lons}
\int_{L_1}^\infty \Phi(L\vert M) dL = 1\,.
\end{equation}
The luminosities  of the satellite  galaxies are also drawn  at random
from $\Phi(L\vert  M)$, but  with the restriction  $L_{\rm min} <  L <
L_1$. 

Note  that   the  resulting   occupation  statistics  are   not  fully
Poissonian.  To  investigate whether such a  deviation from Poissonian
can  be detected from  the statistics  of galaxy  groups we  also, for
comparison, construct a MGRS in which the {\it full} $P(N \vert M)$ is
Poissonian (not  only that  of the satellites),  and in which  all $N$
galaxies are drawn from the  CLF without any restriction other than $L
\geq L_{\rm min}$. In what follows  we refer to the MGRSs based on the
$L_1$-restricted luminosity  sampling as our `fiducial'  mocks, and to
those  with the  unrestricted,  Poissonian sampling  as the  `Poisson'
mocks.

The positions and velocities of  the galaxies with respect to the halo
center-of-mass are drawn assuming that the central galaxy in each halo
resides at rest at the center.  The satellite galaxies follow a number
density  distribution that  is identical  to that  of the  dark matter
particles, and are  assumed to be in isotropic  equilibrium within the
dark matter potential.   To construct MGRSs we use  the same selection
criteria and  observational biases as  in the 2dFGRS,  making detailed
use of  the survey  masks provided by  the 2dFGRS team  (Colless \etal
2001;  Norberg \etal  2002).   Using a  set  of independent  numerical
simulations, we construct 8 independent  MGRSs which we use to address
scatter due to cosmic variance.  The MGRSs thus constructed accurately
match the  clustering properties, the  apparent magnitude distribution
and the  redshift distribution  of the 2dFGRS,  allowing for  a direct
comparison. Finally,  for each MGRS  we construct group  samples using
the same halo-based group finder and the same group selection criteria
as for the 2dFGRS.

Our  fiducial  MGRS, used  throughout  this  paper,  is based  on  the
best-fit CLF listed in Table~1  (the model with ID $\Lambda_{0.9}$) of
van  den   Bosch  \etal  (2004b).    This  CLF  predicts   an  average
mass-to-light ratio on the scale  of clusters of $(M/L)_{\rm cl}=500 h
\MLsun$.   Although in fair  agreement with  independent observational
constraints (e.g.,  Carlberg \etal 1996;  Bahcall \etal 2000,  but see
also  Tully 2003),  we  have  shown that  both  the pairwise  peculiar
velocity dispersions and the group multiplicity function of the 2dFGRS
suggest  a   significantly  higher  cluster   mass-to-light  ratio  of
$(M/L)_{\rm cl}=900 h \MLsun$  (Yang \etal 2004a,b). We therefore also
construct a set  of MGRSs based on the CLF  with $(M/L)_{\rm cl}=900 h
\MLsun$ (see  Yang \etal 2004b),  using the same sampling  strategy as
with our fiducial mocks. Although the model with $(M/L)_{\rm cl}=500 h
\MLsun$  is   preferred  by  the   observed  galaxy-galaxy  clustering
strength,  the  data   is  not  sufficient  to  rule   out  a  cluster
mass-to-light ratio as high as $900 h \MLsun$ (see van den Bosch \etal
2004b).
\begin{table}
\caption{`Mass-limited' group samples from the 2dFGRS}
\begin{tabular}{lccccc}
\hline
Sample & $z_{\rm max}$ & $\log L_{\rm 18,min}$ & $N_{\rm grp}$
              & $\log d$ & $\log M_{\rm min}$\\
       &        & $h^{-2}\Lsun$ &    & $\mpch$   & $h^{-1}\msun$\\
(1)    & (2)    & (3)           & (4)       & (5)    & (6)        \\
\hline\hline
V1 & $0.08$ & $9.4$  & $11682$ & $0.70$ & $11.7$ \\
V2 & $0.09$ & $9.6$  & $13578$ & $0.73$ & $11.8$ \\
V3 & $0.14$ & $10.0$ & $24069$ & $0.83$ & $12.1$ \\
V4 & $0.20$ & $10.5$ & $13510$ & $1.07$ & $12.9$ \\
\hline
\end{tabular}
\medskip
\begin{minipage}{\hssize}
  Column~(1) indicates  the sample ID. The selection  criteria used to
  define  these samples  are indicated  in Columns~(2)  and~(3), which
  list the maximum redshift  $z_{\rm max}$ (sample galaxies obey $0.01
  < z  < z_{\rm max}$) and  the minimum group  luminosity $L_{18, {\rm
      min}}$,  respectively. Columns~(4)  and~(5) list  the  number of
  groups in each sample, $N_{\rm grp}$, and the mean group separation,
  $d$,  respectively.   Finally,  column~(6) lists  the  corresponding
  minimum  halo  mass,  $M_{\rm  min}$, obtained  using  the  relation
  between   $d$   and   $M$   shown   in  the   left-hand   panel   of
  Fig.~\ref{fig:M_L_d} (assuming $\sigma_8=0.9$).
\end{minipage}
\end{table}

\subsection{Ranking halo mass according to group luminosity}
\label{sec:ranking}

In order to infer halo occupation statistics from our group samples it
is crucial  that we can estimate  the halo masses  associated with the
groups. For individual, rich  clusters one could in principle estimate
halo masses using the kinematics of the member galaxies, gravitational
lensing of background sources, or the temperature profile of the X-ray
emitting gas.   For most groups,  however, no X-ray emission  has been
detected, and  no lensing  data is available.   In addition,  the vast
majority  of the  groups in  our sample  contain only  a  few members,
making  a  dynamical mass  estimate  based  on  its members  extremely
unreliable.  We  thus need to  adopt a different approach  to estimate
halo masses.

As discussed in  YMBJ, for each group one  can define a characteristic
luminosity,  $L_{18}$, defined as  the total  luminosity of  all group
members  brighter  than  $M_{b_J}-5\log  h  =  -18$.   For  groups  at
relatively high redshift $L_{18}$ can not be measured directly because
of the  apparent magnitude  limit of the  survey. For these  groups we
estimate $L_{18}$ from the luminosity of the observable group members,
using a  correction factor that is calibrated  using relatively nearby
groups (see Yang  \etal (2004b,c) for details). Tests  with MGRSs have
shown that  $L_{18}$ is tightly correlated  with the mass  of the dark
matter  halo hosting  the  group.   As shown  in  Yang \etal  (2004c),
ranking  groups   according  to  $L_{18}$  is   therefore  similar  to
mass-ranking,  allowing the  construction of  reliable, `mass-limited'
group samples.   In Table~1, we list the  `mass-limited' group samples
used  in  this paper.   Each  sample  is  specified by  two  selection
criteria;  a  lower  limit  on  $L_{18}$, which  as  we  argued  above
translates into  a lower  limit on halo  mass, and a  maximum redshift
$z_{\rm max}$.   The latter  assures that each  sample is  complete to
some  absolute magnitude  limit, which  is required  for  a meaningful
comparison of the group member galaxies.

In  order to convert  the $L_{18}$-ranking  to the  corresponding halo
mass, $M$, we  use the mean group separation,  $d=n^{-1/3}$, as a mass
indicator. Here $n$  is the number density of  all groups brighter (in
terms of $L_{18}$) than the group in consideration.  In the left panel
of Fig.~\ref{fig:M_L_d},  we plot the mean relation  between the group
luminosity $L_{18}$ and the mean  group separation $d$ for 2dF groups,
with  different   lines  corresponding  to   different  `mass-limited'
subsamples. Overall the $L_{18}$-$d$ relation is similar for different
subsamples.   The small,  but noticeable,  differences  reflect cosmic
variance due  to the presence  of a few  very large structures  in the
2dFGRS  (see  e.g., Baugh  \etal  2004).   Since  $L_{18}$ is  tightly
correlated with halo mass, we  can convert $d$ to $M$.  Unfortunately,
this conversion requires knowledge of the halo mass function, and thus
knowledge   of   the  cosmological   parameters.    As  discussed   in
Section~\ref{sec:mock},   throughout   this   paper  we   consider   a
$\Lambda$CDM concordance cosmology  with $\sigma_8=0.9$. To illustrate
how  sensitive  the  $d$-to-$M$   conversion  depends  on  the  rather
uncertain power-spectrum normalization parameter, the right-hand panel
of   Fig.~\ref{fig:M_L_d}  plots  the   $M$-$d$  relations   for  both
$\sigma_8=0.9$  and $\sigma_8=0.7$.   For haloes  with masses  $M \lta
10^{13.5} h^{-1} \msun$, the $M$-$d$ relation is virtually independent
of $\sigma_8$.  For  $M \gta 10^{13.5} h^{-1} \msun$,  however, $d$ is
smaller in the $\sigma_8=0.9$  cosmology simply because massive haloes
are  more  abundant in  cosmologies  with  larger $\sigma_8$.   Unless
specifically  stated otherwise,  we  use the  $\sigma_8=0.9$ model  to
convert $d$ to  $M$, but we emphasize that any function  of $M$ can be
converted back into  a function of $d$ using  the relation represented
by the solid curve in the right-hand panel of Fig.~\ref{fig:M_L_d}.
\begin{figure*}
\centerline{\psfig{figure=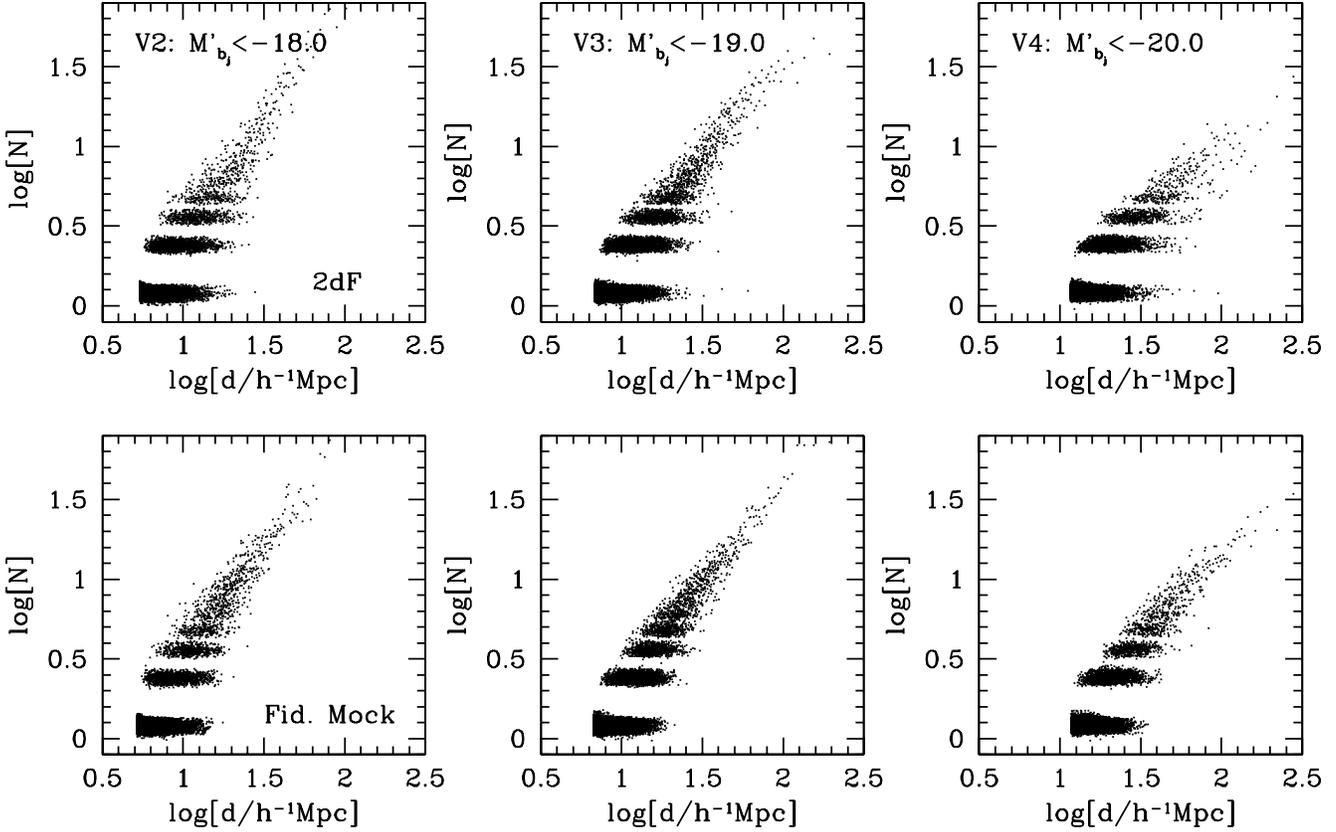,width=\hdsize}} 
\caption{The halo occupation number distributions for groups in the 
  2dFGRS  (upper panels)  and  one realization  of  the fiducial  MGRS
  (lower panels) as a function  of the group mean separation $d$.  The
  left-,  middle   and  right-hand  panels   correspond  to  different
  `mass-limited'   samples  with  an   absolute  magnitude   limit  of
  $M'_{b_J}=M_{b_J}-5\log  h$.  Note that  the occupation  numbers are
  corrected   for  incompleteness   effects,   which  explains   their
  non-integer nature.}
\label{fig:2dF_HO}
\end{figure*}
\begin{figure*}
\centerline{\psfig{figure=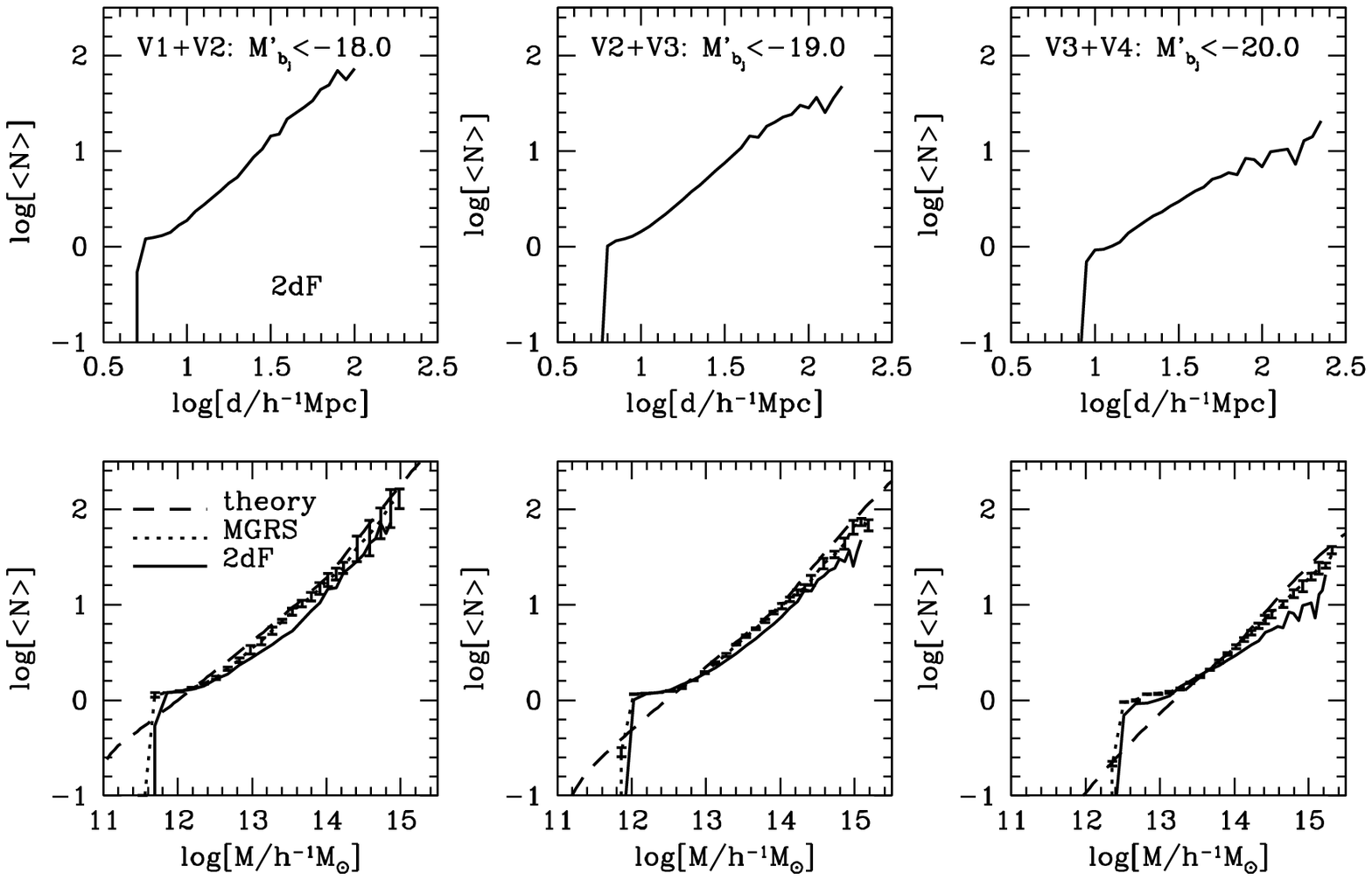,width=\hdsize}} 
\caption{The  mean  halo occupation numbers as a function of  mean  
  group separation  $d$ (upper panels)  and halo mass (lower  panels). 
  Solid lines  correspond to groups  in the 2dFGRS.  The  dotted lines
  with errorbars  in the lower  panels show the results  obtained from
  the groups in  our fiducial MGRSs.  The errorbars  are obtained from
  the 1-$\sigma$ scatter among  8 independent MGRSs.  The dashed lines
  in  the lower  panels, labelled  `theory', indicate  the  true, mean
  occupation    numbers,    obtained    directly    from    the    CLF
  (eq.~[\ref{averN}]) used  to construct the MGRS.   A comparison with
  the dotted lines  shows that  the halo  occupation numbers  are well
  recovered, except for groups with  $\langle N \rangle < 1$, where an
  artificial shoulder and break are introduced. The comparison between
  2dFGRS  and MGRS  shows that  our fiducial  model predicts  too many
  galaxies per  group at the  high mass end  (see text for  a detailed
  discussion).}
\label{fig:2dF_HO2}
\end{figure*}

\section{Halo occupation statistics from 2dFGRS groups}
\label{sec:property}

Having assigned 2dFGRS galaxies  into groups according to their common
dark matter  haloes, we  now present a  detailed investigation  of the
halo  occupation statistics,  describing how  galaxies  with different
physical  properties  are  associated   with  dark  matter  haloes  of
different mass.

\subsection{The halo occupation distribution}
\label{sec:hod}

The upper  three panels  of Fig.~\ref{fig:2dF_HO} plot  the occupation
numbers of galaxies  in 2dFGRS groups as a function  of $d$ (which, as
discussed above, can  be used as a proxy for  halo mass).  Results are
shown  for  galaxies with  $M_{b_J}-5\log  h  <  -18.0$ (left  panel),
$-19.0$ (middle panel) and  $-20.0$ (right panel), respectively. These
occupation numbers  are obtained  using the summation  $N=\sum 1/c_i$,
with $c_i$  the completeness in the  2dFGRS at the  position of galaxy
$i$,  so that  $N$ is  not necessarily  an integer.   The  lower three
panels of  Fig.~\ref{fig:2dF_HO} plot the same  occupation numbers but
this  time obtained  from our  fiducial MGRS,  using exactly  the same
method as for  the 2dFGRS.  Note that in the  construction of the MGRS
we use  completeness maps  of the 2dFGRS.   Therefore, we  can compute
exactly the same $N$ for our mock groups (i.e., summation of $c_i$) as
in the real 2dFGRS.  Although  the occupation statistics of the groups
in the MGRS  reveal overall the same behavior as  those in the 2dFGRS,
there  are some  noticeable  differences, which  we  quantify in  more
detail below.

The upper  panels of Fig.~\ref{fig:2dF_HO2}  plot the {\it  mean} halo
occupation numbers, $\langle N \rangle$,  as a function of $d$ for the
same samples of 2dFGRS  groups as in Fig.~\ref{fig:2dF_HO}.  Using the
$M$-$d$ relations shown in  Fig.~\ref{fig:M_L_d} we convert these into
the average occupation  numbers as function of halo  mass shown in the
bottom panels. Note that  the average occupation numbers increase with
halo mass, as  expected.  At the low mass end,  however, they reveal a
relatively flat shoulder and a sharp break, both at $\langle N \rangle
\sim 1$.  This  sharp break seems to indicate  an almost deterministic
relation between  the luminosity of the central  (brightest) galaxy in
each halo and  the mass of the halo, while  the shoulder suggests that
the  second  brightest  galaxy   is  significantly  fainter  than  the
brightest  one  (e.g., Zheng  \etal  2004).   The  dotted curves  with
errorbars indicate the occupation  numbers obtained from the groups in
our  fiducial MGRS.  These  reveal an  almost identical  shoulder plus
break. The  dashed lines, however, indicate the  $\langle N \rangle_M$
obtained directly  from the CLF  used to construct the  MGRS (computed
from eq.~[\ref{averN}] with $L_{\rm min}$ the minimum luminosity of the
sample  under  consideration).  The  agreement  of  these  {\it  true}
occupation  statistics  with  those   inferred  from  the  MGRS  group
catalogues is remarkably good at $\langle N\rangle \gta 1$, indicating
that the conversion  of $L_{18}$ to halo mass  via the mean separation
$d$ does not  introduce any systematic error.  It  also shows that the
groups identified with the method  developed in Yang \etal (2004b) can
be used to accurately  probe halo occupation statistics.  For $\langle
N \rangle \lta 1$, however,  there is a noticeable discrepancy between
the true  $\langle N \rangle_M$ and  that obtained from  the MGRS.  In
particular, the shoulder and sharp  break at $\langle N \rangle \simeq
1$ visible in  the mean occupation numbers of the  mock groups are not
present  in  the true  $\langle  N  \rangle_M$.   The origin  of  this
discrepancy is  easy to understand  if one takes the  stochasticity of
the occupation  numbers into account.   Since we estimate  halo masses
from  the  $L_{18}$-ranking, halo  masses  are  overestimated if  they
happen to contain  a relatively bright galaxy (compared  to the mean). 
Similarly, haloes with a relatively  faint galaxy compared to the mean
will have  their masses underestimated.  If the  average luminosity is
close to the luminosity limit of the sample, this stochasticity in the
occupation  statistics causes  a  systematic deviation  from the  true
$\langle  N  \rangle_M$,  as  the  haloes with  the  relatively  faint
galaxies  will not  make the  sample selection  criteria.   Caution is
therefore required  in interpreting the  sharp break and  the shoulder
around $N=1$ seen  in the occupation statistics of  the 2dFGRS groups. 
Although they  may still  be real,  we cannot rule  out that  they are
simply artifacts due to combined effect of stochasticity and magnitude
limit.
\begin{figure*}
\centerline{\psfig{figure=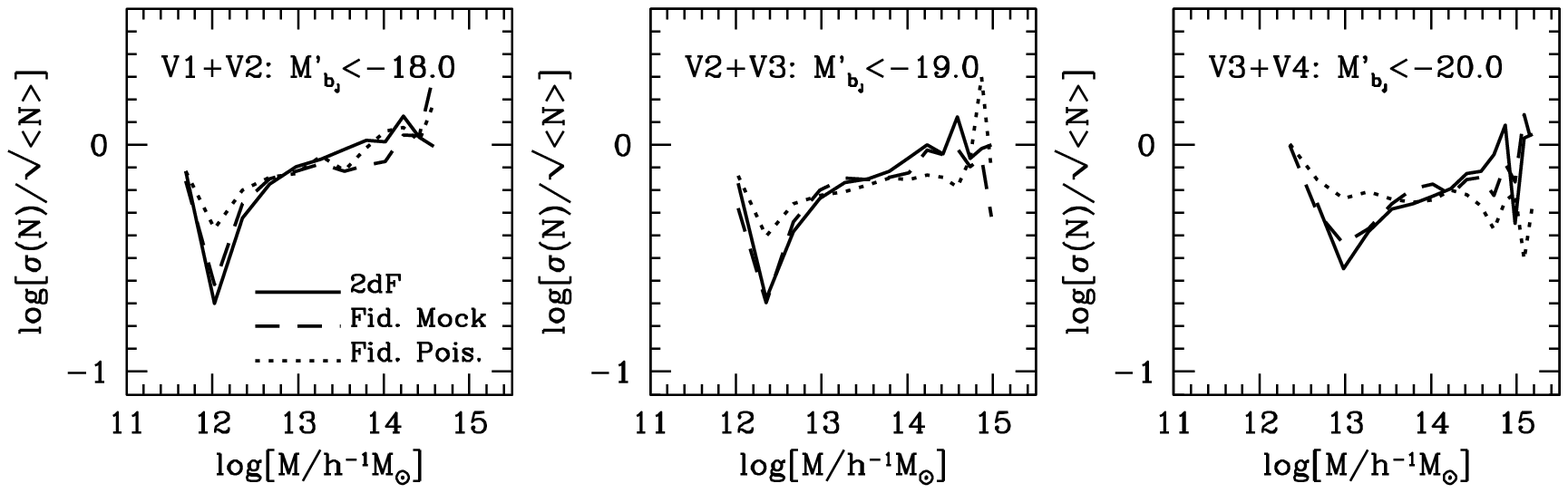,width=\hdsize}} 
\caption{The scatter of the halo occupation number distribution,
  expressed  in terms  of the  ratio between  the  standard deviation,
  $\sigma(N)$,  and the  square  root of  the  mean, $\sqrt{\langle  N
    \rangle}$. Note  that this ratio is  equal to unity  for a Poisson
  distribution.  Results  are shown as  function of halo mass  $M$ for
  the same  `mass-limited' groups samples  as in Figs~\ref{fig:2dF_HO}
  and~\ref{fig:2dF_HO2}.  The solid and dashed lines correspond to the
  results  obtained from  the groups  in the  2dFGRS and  the fiducial
  MGRS, respectively, and are in  excellent agreement with each other. 
  The  dotted  lines correspond  to  a MGRS  that  is  similar to  the
  fiducial one,  except that the  luminosity of the central  galaxy is
  not treated in  a special way (i.e., the  true occupation statistics
  are  purely  Poissonian  in  this  case).   Since  halo  masses  are
  estimated  from the  ranking of  $L_{18}$, the  ratio  deviates from
  unity even for  this pure Poisson case. See the  text for a detailed
  discussion regarding the interpretation of these results.}
\label{fig:f_d}
\end{figure*}

The lower panels of Fig.~\ref{fig:2dF_HO2} also show that our fiducial
MGRSs predict too many galaxies per  group at the high mass end. Given
the errorbars, which reflect the scatter among 8 fiducial MGRSs, these
differences    are    very   significant.     As    we   discuss    in
Section~\ref{sec:central}, this reflects a  problem with the number of
satellite galaxies, and suggests  either a high mass-to-light ratio on
the scale  of galaxy  clusters, or a  reduction of  the power-spectrum
normalization  $\sigma_8$ from the  fiducial value  of $0.9$  to $\sim
0.7$.  This is easy  to understand: increasing the mass-to-light ratio
on  the  scale  of  clusters,  basically implies  fewer  galaxies  per
cluster.   Since the  total number  density of  galaxies  is conserved
(constrained by  the galaxy  luminosity function), these  galaxies now
need to be distributed  over lower mass haloes.  Therefore, increasing
$(M/L)_{\rm cl}$  decreases $\langle N  \rangle$ for the  most massive
haloes,  while (mildly) increasing  $\langle N  \rangle$ for  the less
massive haloes. Note that, since less massive haloes are less strongly
clustered,  an  increase  of   $(M/L)_{\rm  cl}$  lowers  the  overall
clustering strength  of the galaxy  population.  However, as  shown in
van den  Bosch, Mo \&  Yang (2003b) and  van den Bosch  \etal (2004a),
there is  a sufficient amount  of freedom in  the data to allow  us to
modify $(M/L)_{\rm cl}$ and still get  a reasonable match to the data. 
It is this freedom  that we exploit here to argue for  a high value of
$(M/L)_{\rm cl}$. An alternative  solution to the mismatch between the
$\langle  N \rangle_M$  of MGRS  and 2dFGRS  is to  lower  $\sigma_8$. 
Lowering $\sigma_8$ reduces  the number of massive haloes, but 
changes little the clustering strength of haloes at a given mass
(the decrease in the clustering strength of dark matter particles
is largely compensated by the increase in the bias factor).  
If the value of $(M/L)_{\rm cl}$ is fixed, lowering $\sigma_8$ 
requires more galaxies to be assigned to lower-mass haloes,
and the net effect on galaxy clustering is similar to that with 
a higher value of $(M/L)_{\rm cl}$ 
(Yang \etal 2004a; van den Bosch \etal 2004b).

In addition to the {\it  mean} occupation numbers, we also investigate
the second moment of  the halo occupation distribution.  This quantity
is required in the modelling  of the two-point correlation function of
galaxies on small scales (e.g., Benson \etal 2000; Berlind \etal 2003;
Yang  \etal  2004a), and  holds  important  information regarding  the
physical   processes  related   to  galaxy   formation.    In  earlier
investigations, a number of simple models were adopted to describe the
second moment  of the halo occupation distribution  and its dependence
on  halo  mass (e.g.,  Berlind  \&  Weinberg  2002).  With  our  group
samples, we  can actually measure this quantity  directly.  We present
our  results in  terms of  the ratio  between the  standard deviation,
$\sigma(N)$,  and  the  square  root  of the  mean,  $\sqrt{\langle  N
  \rangle}$.    Since   for   a   Poisson   distribution   $\sigma(N)/
\sqrt{\langle  N \rangle}  = 1$,  this ratio  expresses the  amount of
stochasticity   relative    to   Poisson.    The    solid   lines   in
Fig.~\ref{fig:f_d} show  the results  obtained from the  2dFGRS, where
the three panels  correspond to the same volume-limited  samples as in
Figs.~\ref{fig:2dF_HO}      and~\ref{fig:2dF_HO2}.       The     ratio
$\sigma(N)/\sqrt{\langle  N \rangle}$  is  close to  unity in  massive
haloes, but  reveals a pronounced  minimum at low $M$.   This suggests
that  the halo  occupation distribution  is (close  to)  Poissonian in
massive haloes and significantly sub-Poissonian in low mass haloes.

Note,  however,  that  because  of  our method  of  assigning  masses,
Fig.~\ref{fig:f_d} really shows the scatter  in $N$ at given $L_{18}$. 
In  order to  test how  the scatter  in the  relation between  $M$ and
$L_{18}$ impacts on these results,  we compare our findings with those
obtained from  a MGRS that is  identical to our  fiducial MGRS, except
that this time  the central galaxy is not treated  in any special way,
so  that the occupation  distribution, $P(N  \vert M)$,  is completely
Poissonian (see Section~\ref{sec:mock}).   Any deviation of $\sigma(N)
/\sqrt{\langle  N\rangle}$  from  unity  in  this  MGRS  is  therefore
completely  artificial, allowing us  to assess  the robustness  of our
findings.  The  dotted curves  in Fig.~\ref{fig:f_d} show  the results
obtained   from  this   MGRS.   They   reveal  a   small   minimum  in
$\sigma(N)/\sqrt{\langle N \rangle}$ at small $M$, similar though less
pronounced  than for  the  2dFGRS.   The origin  of  this artefact  is
similar to that of the artificial shoulder and break in the {\it mean}
occupation numbers.  In  haloes with $\langle N \rangle  \simeq 1$ one
expects a significant fraction of haloes  with $N = 0$; in fact, for a
Poissonian $P(N \vert  M)$ the probability to have  $N=0$ is almost 40
percent.  These  haloes, however, do  not appear in the  group samples
causing an overestimate of $\langle N \rangle$ and an underestimate of
the  variance.    Therefore,  the  ratio   $\sigma(N)/\sqrt{\langle  N
  \rangle}$  is  underestimated for  haloes  with  $\langle N  \rangle
\simeq  1$.   The upturn  at  the  very low-mass  end  is  due to  the
(artificial)  sharp  break  in   $\langle  N  \rangle$,  which  drives
$\sigma(N)/\sqrt{\langle N \rangle}$ up  again.  The presence of these
artefacts clearly demonstrates the  importance of using detailed MGRSs
to properly interpret the data.
\begin{figure*}
\centerline{\psfig{figure=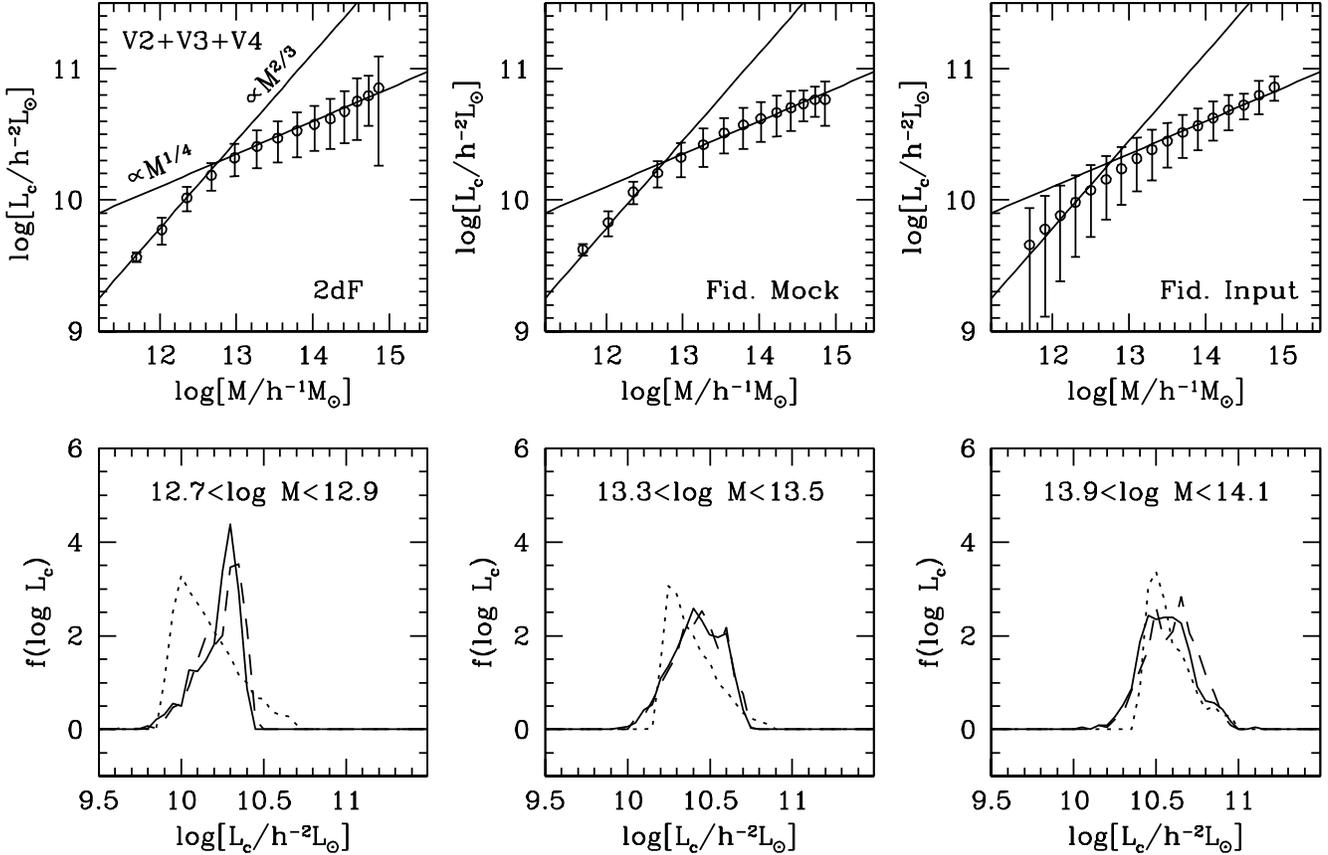,width=\hdsize}} 
\caption{The upper  panels plot the mean central galaxy luminosity,
  $L_c$, as function of halo  mass, $M$, with the errorbars indicating
  the 1-$\sigma$  scatter around the  mean. The left and  middle panel
  correspond to  the 2dFGRS and the fiducial  MGRS, respectively, where
  we  have  used  the  `mass-limited'  samples V2,  V3,  and  V4  (see
  Table~1).   Solid lines  indicate two  power-law relations,  and are
  indicated to  facilitate a comparison. Note  the excellent agreement
  between  the  2dFGRS  and  the  MGRS.  The  upper  right-hand  panel
  indicates the  same relation  between $L_c$ and  $M$, but  this time
  determined directly from the  populated haloes in our simulation box
  (i.e., without making  a mock redshift surveys from  which we select
  groups). Although  the mean $L_c-M$ relation  is virtually identical
  to that derived  from the mock groups, the  scatter is significantly
  larger  in low  mass haloes  (see  text for  discussion). The  lower
  panels plot  the distributions $P(L_c \vert M)$  for three different
  bins in  halo mass, as  indicated. Solid, dashed, and  dotted curves
  correspond to the 2dFGRS, the fiducial MGRS, and the simulation box,
  respectively.}
\label{fig:2dF_Lc}
\end{figure*}

The dashed lines in  Fig.~\ref{fig:f_d} show the results obtained from
our  {\it fiducial} MGRS.   These results  are in  excellent agreement
with those obtained  from the 2dFGRS, indicating that  the CLF and the
method used for its sampling agree well with the data. As indicated in
Section~\ref{sec:mock},  the  occupation  statistics  of  the  central
galaxies are treated differently than those of the satellite galaxies:
whereas $P(N \vert M)$ is  Poissonian for the latter, central galaxies
follow a much narrower  nearest-integral distribution. This means that
$P(N \vert  M)$ is strongly sub-Poissian whenever  $\langle N \rangle$
is small. As is evident from a comparison with Fig.~\ref{fig:2dF_HO2},
the minimum  in $\sigma(N) /\sqrt{\langle N\rangle}$ occurs  at a halo
mass  where the average  occupation number  is virtually  unity. This,
together with  the fact that the minimum  in $\sigma(N) /\sqrt{\langle
  N\rangle}$ is much more pronounced than in the pure-Poissonian MGRS,
leads us to conclude that (i) the number of satellite galaxies above a
certain luminosity  limit follow  a Poissonian distribution,  and (ii)
the  occupation  statistics of  central  galaxies are  sub-Poissonian,
indicating some deterministic  behavior in galaxy formation.  Clearly,
a  detailed  study  of  the  higher-order moments  of  the  occupation
statistics can yield important constraints on galaxy formation, and we
intend to return to this in more detail in a forthcoming paper.
\begin{figure}
\centerline{\psfig{figure=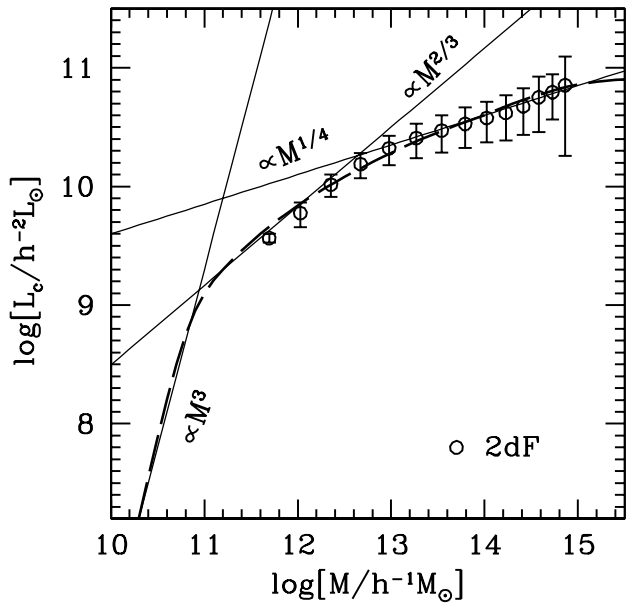,width=\hssize}} 
\caption{The mean central galaxy luminosity, $L_c$, as function of
  halo  mass, $M$,  over a  large range  in haloes  masses.   The data
  points  are the  same as  those  shown in  the upper  left panel  of
  Fig~\ref{fig:2dF_Lc}.   The dashed curve  is the  $L_c$-$M$ relation
  given  by the  CLF obtained  from matching  the  observed luminosity
  function of  galaxies and  the correlation length  as a  function of
  galaxy  luminosity.  Note  the existence  of  another characteristic
  mass  scale, $M\simeq  10^{11}\msunh$, below  which  $L_c$ decreases
  rapidly  with  decreasing $M$.   This  plot  indicates that  scaling
  relations such as the Tully-Fisher relation hold only over a limited
  range of halo masses.  }
\label{fig:2dF_Lc2}
\end{figure}

\subsection{Central versus satellite galaxies}
\label{sec:central}

In  theoretical models of  galaxy formation,  galaxies in  dark matter
haloes  are  usually separated  into  central  galaxies and  satellite
galaxies. Since  central and satellite  galaxies are expected  to have
somewhat  different  formation histories  (e.g.,  Kauffmann, White  \&
Guiderdoni  1993), it  is  interesting to  study  the halo  occupation
distribution  separately for  these  two categories  of galaxies.   By
definition, the  central galaxy in  a halo should  be the one  that is
located near the center of the  host halo. Since in theory the central
galaxy is  expected to be the  most massive one among  all galaxies in
the halo,  we have defined the  brightest galaxy in a  group (halo) as
the central galaxy, and the others as satellite galaxies.

The upper  panels of  Fig.~\ref{fig:2dF_Lc} plot the  relation between
the luminosity of the central galaxy,  $L_c$, and the mass of the host
halo,  $M$.  Results are  shown both  for groups  in the  2dFGRS (left
panel) and for those in our fiducial MGRS (middle panel). We also show
the  true  relation  between  $L_c$  and $M$  (right  panel)  obtained
directly  from  the populated  haloes  in  our  simulation box  (i.e.,
without making a  mock redshift surveys from which  we select groups). 
The  mean  $L_c$-$M$ relation  is  remarkably  similar  for all  three
samples,  and well described  by a  broken power-law  with $L_c\propto
M^{2/3}$ at $M\la 10^{13} h^{-1}M_{\odot}$ and $L_c\propto M^{1/4}$ at
$M\ga  10^{13} h^{-1}M_{\odot}$.   At  the low-mass  end,  this is  in
excellent agreement with results  based on galaxy-galaxy weak lensing,
which imply  that $M\propto L_c^{1.5}$ (e.g.  Yang  \etal 2003a; Guzik
\& Seljak 2002).  At the massive end, $L_c$ only increases very slowly
with halo mass, which is consistent with the recent result obtained by
Lin \etal  (2004), indicating  that there must  be a  physical process
that prevents  the central  galaxies in massive  haloes from  growing. 
One  possibility  is  that  radiative  cooling  of  halo  gas  becomes
negligible in massive haloes,  with $M \simeq 10^{13} h^{-1}M_{\odot}$
the characteristic  mass that marks  the transition from  effective to
ineffective  cooling (cf., Dekel  2004).  The  requirement for  such a
transition  is  well  known  from  semi-analytical  models  of  galaxy
formation  where it is  required to  reproduce the  bright end  of the
observed  luminosity function  (e.g.,  White \&  Rees 1978;  Kauffmann
\etal 1993; Benson \etal 2003; Kang \etal 2004).

The errorbars  in the  upper panels of  Fig.~\ref{fig:2dF_Lc} indicate
the  scatter  around  $L_c$  at  given $M$.   Except  for  some  small
discrepancies at  high $M$, the amounts  of scatter in  the 2dFGRS and
MGRS are very similar. This  is illustrated more clearly in the lower
panels of  Fig.~\ref{fig:2dF_Lc}, which plot  the actual distributions
of  $L_c$ for  three bins  in halo  mass (as  indicated) for  both the
2dFGRS  (solid  lines)  and  the  MGRS (dashed  lines).   Overall  the
agreement is remarkably good, providing strong support for the CLF and
its sampling strategy. Note that  the $P(L_c \vert M)$ look similar to
log-normal distributions, with a fairly narrow width that depends only
mildly on halo mass.

To properly interpret these findings we compare these $P(L_c \vert M)$
with  those  obtained  directly  from  the  populated  haloes  in  the
simulation  box (i.e.,  without making  a mock  redshift  surveys from
which we  select groups).   As evident from  the upper-right  panel of
Fig.~\ref{fig:2dF_Lc},  at  low $M$  the  scatter  in  the {\it  true}
$L_c-M$ relation is much larger than in the relation inferred from the
mock group catalogue.   This is also evident from  the lower panels in
Fig.~\ref{fig:2dF_Lc}  which  show  that  the true  $P(L_c  \vert  M)$
(dotted  curves) in  haloes  with  $M \lta  10^{13}  h^{-1} \Msun$  is
significantly broader  than the inferred  distribution. In particular,
the inferred  distribution seems  to lack predominantly  the low-$L_c$
galaxies.   This  discrepancy arises  from  the  stochasticity in  the
$L_{18}-M$ relation. In low-mass  haloes, where the average occupation
number is  close to unity, $L_{18}$  is basically identical  to $L_c$. 
This means that the $L_{18}$-ranking becomes similar to $L_c$-ranking,
so that the resulting $L_c-M$ relation becomes virtually scatter free.
In addition, because  of the magnitude limit of  the group sample, the
haloes with  relatively faint central  galaxies are missed,  causing a
deficit  of  low-$L_c$ galaxies.   On  the  other  hand, some  central
galaxies  may  be  missed  in  the  sample  because  of  observational
selection  effects, and  so some  of  the galaxies  identified as  the
central galaxies are  actually the second or even  the third brightest
galaxy  in a  group. This  introduces  extra scatter  in $L_c$,  which
causes the  scatter at the  high-mass end to  be larger for  the MGRSs
(and the 2dFGRS) than the  {\it true} scatter.  Therefore, as with the
scatter  in  the  occupation  numbers,  great care  is  required  when
interpreting  the scatter  in $P(L_c  \vert M)$.   In  particular, the
log-normal character of  $P(L_c \vert M)$ of the  2dFGRS galaxies does
not necessarily  imply that the  true distribution is  log-normal.  We
emphasize, however, that despite this bias, the comparison between the
2dFGRS  and the  MGRS is  still meaningful.   In particular,  the good
agreement between  both group catalogues  suggests that our  method of
assigning galaxies to  dark matter haloes used in  the construction of
the MGRS  (see Section~\ref{sec:mock}) is in  excellent agreement with
the 2dFGRS.

Since our group  sample becomes quite incomplete for  halos with $M\la
10^{12}h^{-1}{\rm  M}_\odot$, we cannot  use our  groups to  study the
$L_c$-$M$ relation  for low-mass haloes.   However, our CLF,  which is
constrained by the abundances  and clustering properties of the galaxy
population, {\it does} contain  such information.  The dashed curve in
Fig.~\ref{fig:2dF_Lc2} shows the  $L_c$-$M$ relation obtained from our
CLF down to haloes with $M = 10^{10} h^{-1} \Msun$. For comparison, we
also plot the results obtained  from our 2dFGRS group catalogue, which
are  in excellent agreement  with these  predictions. For  haloes with
$M\la 10^{12}h^{-1}{\rm M}_\odot$,  however, no reliable determination
of $L_c(M)$ can  be obtained from the groups.   This is unfortunate as
our CLF  model predicts the  presence of a second  characteristic mass
scale at  $M\simeq 10^{11}\msunh$.  For haloes below  this scale $L_c$
is  predicted to  decrease  rapidly with  decreasing  $M$ (roughly  as
$L_c\propto M^{\beta}$,  with $\beta\sim 2$ - $4$).   This is required
in order to match the relatively steep slope of the halo mass function
at the low mass end with the relatively shallow faint-end slope of the
galaxy luminosity function (see e.g.,  Yang \etal 2003b), and is often
interpreted  as  `evidence' for  a  suppression  of  star formation  by
feedback effects (e.g. Dekel \&  Silk 1986; Dekel 2004). Note that the
$L_c$-$M$  relation  shown  in  Fig.~\ref{fig:2dF_Lc2}  suggests  that
scaling relations such as the Tully-Fisher relation can hold only over
a limited range of halo masses.

\begin{figure*}
\centerline{\psfig{figure=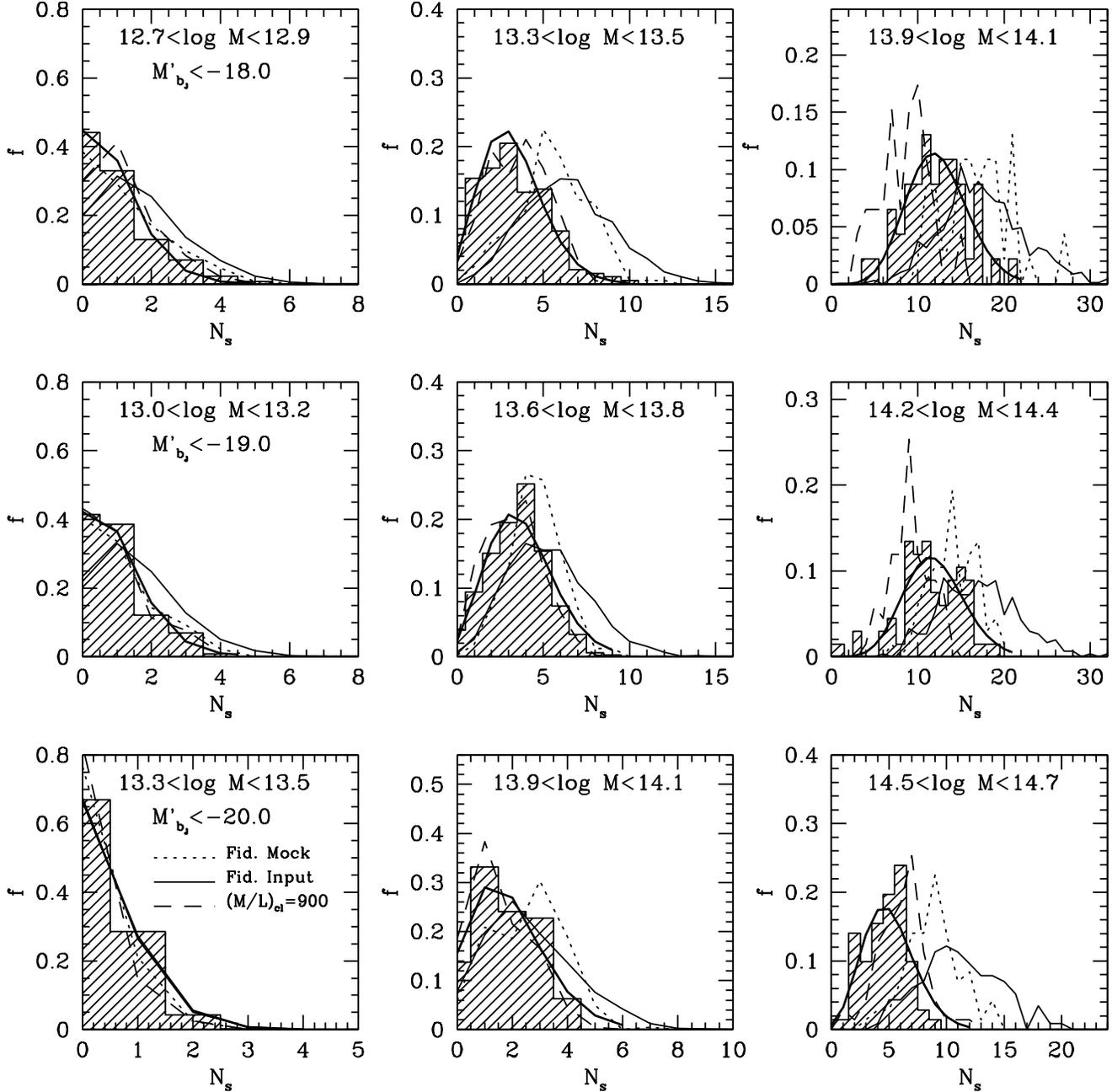,width=\hdsize}} 
\caption{Distributions of the number of satellite galaxies in groups
  for different bins in halo  mass, as indicated. Panels in the upper,
  middle  and lower  rows correspond  to different  absolute magnitude
  limits as  indicated, where $M'_{b_J}=M_{b_J}-5\log  h$.  The hatched
  histograms indicate  the distributions  obtained from the  groups in
  the 2dFGRS.  Thick solid  curves correspond to Poisson distributions
  with the same mean $N_s$, and are shown to illustrate the Poissonian
  nature  of  $P(N_s \vert  M)$.   The  dotted  and dashed  histograms
  indicate the  distributions obtained from the fiducial  MGRS and the
  MGRS with $(M/L)_{\rm cl} = 900 h \MLsun$, respectively. Whereas the
  former  dramatically overestimates the  average number  of satellite
  galaxies  in massive  haloes,  the latter  fits  the 2dFGRS  results
  extremely well. The thin,  solid lines, are the number distributions
  of satellite galaxies obtained directly from the populated haloes
  in our fiducial simulation box (i.e., without making a mock redshift
  surveys from  which we select  groups). These therefore  reflect the
  true $P(N_s  \vert M)$.  Note the good,  overall agreement  with the
  distributions obtained from the fiducial MGRS (dotted curves).}
\label{fig:f_Ns}
\end{figure*}

Let us  now move on to satellite  galaxies.  Fig.~\ref{fig:f_Ns} plots
the  distribution  of the  number  of  satellite  galaxies in  groups,
$N_{\rm s}$,  for a  number of different  mass bins.  The  thick solid
curves indicate Poisson distributions with the same $\langle N_{\rm s}
\rangle$, and fit the  $N_{\rm s}$-distributions extremely well.  This
is  an important  result, because  it suggests  a direct  link between
satellite  galaxies and  dark matter  subhaloes.  In  a  recent study,
Kravtsov \etal  (2004), using large numerical  simulations, have shown
that the  occupation distribution of  dark matter subhaloes  follows a
Poisson  distribution,  in  excellent  agreement with  the  occupation
statistics of the satellite galaxies shown here.

For  comparison,  the dotted  lines  in  Fig.~\ref{fig:f_Ns} plot  the
distributions of $N_{\rm s}$  obtained from our fiducial MGRS.  Unlike
with the central galaxies, for  which MGRS and 2dFGRS are in excellent
agreement,  the  MGRS contains  far  too  many  satellite galaxies  in
massive  systems.   This  explains  the  discrepancy  in  the  average
occupation  numbers  $\langle  N  \rangle$  at large  $M$  shown  in
Fig.~\ref{fig:2dF_HO2}, and is consistent with the findings in some of
our  previous studies  (Yang \etal  2004a; YMBJ;  van den  Bosch \etal
2004b).  As discussed  in Yang \etal (2004a), there  are two different
ways to  reduce the number  of rich systems.   One is to  increase the
mass-to-light  ratio  of clusters,  so  that  the  number of  galaxies
assigned to  a massive  halo is  reduced. The other  is to  reduce the
value of  $\sigma_8$, so  that the number  of massive haloes  that can
host  a  large  number  of  satellite  galaxies  is  reduced.   As  an
illustration, the dashed lines in Fig.~\ref{fig:f_Ns} show the results
obtained from a MGRS based on the CLF model with $(M/L)_{\rm cl}=900 h
\MLsun$  (see Section~\ref{sec:mock}). This  model matches  the 2dFGRS
distributions much better.  Similar  results are obtained if we reduce
the value of  $\sigma_8$ to $\sim 0.70$ (not  shown).  Note that these
two models are also favored by several other observations based on the
2dFGRS, such as the  redshift-space clustering of galaxies (Yang \etal
2004a)  and the  multiplicity function  of galaxy  groups  (Yang \etal
2004b).

Before concluding  that therefore the  satellite occupation statistics
hint towards either a high $(M/L)_{\rm cl}$ or a low $\sigma_8$, it is
important to  check whether  the $L_{18}$ to  $M$ conversion  (via the
mean separation  $d$) used,  has not introduced  any artifact  in this
statistic.  To test this we  determine the distribution of $N_{\rm s}$
directly  from  the populated  haloes  in  the  simulation box  (i.e.,
without making a  mock redshift surveys from which  we select groups). 
Since we know the halo mass exactly for each halo in the box,  we can
compute  the number of  satellite galaxies  above the  magnitude limit
listed.   The resulting  distributions  of $N_{\rm  s}$  are shown  in
Fig.~\ref{fig:f_Ns}  as  thin  solid   lines  and  are  in  reasonable
agreement with the distributions  obtained from the corresponding MGRS
(dotted lines).  Although small  differences are apparent, the overall
trends, especially  the dramatic overprediction of  the mean satellite
number, is nicely reproduced.   This demonstrates that the discrepancy
between  2dFGRS  and MGRS  is  real,  indicating  that either  cluster
mass-to-light ratios are high or that $\sigma_8 \sim 0.7$.
\begin{figure*}
\centerline{\psfig{figure=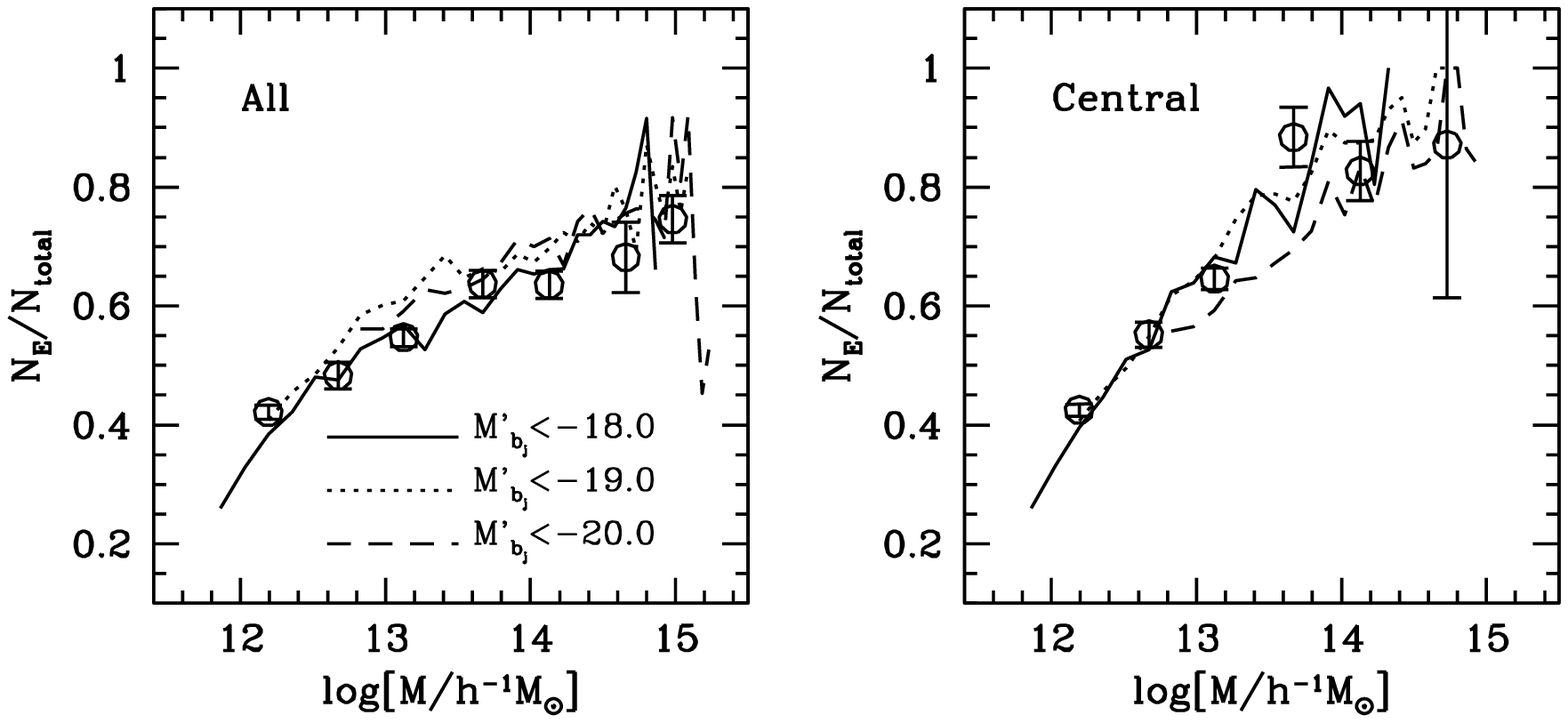,width=\hdsize}} 
\caption{{\it Left-hand panel:} The fraction of early-type galaxies
  in groups of the 2dFGRS as  function of halo mass. Solid, dotted and
  dashed lines correspond to  the `mass-limited' group samples V2, V3,
  and V4  with different absolute magnitude  limits, respectively (see
  Table~1). As  a comparison, the  results for groups in  the fiducial
  MGRSs  with $M_{b_J}-5\log  h <  -18.0$  are shown  as circles  with
  errorbars  (1-$\sigma$  scatter   among  8  fiducial  MGRSs).   {\it
    Right-hand panel:}  Same as left-hand panel, except  that now only
  central galaxies are considered. }
\label{fig:types}
\end{figure*}

\subsection{Dependence on galaxy type}
\label{sec:types}

Madgwick \etal  (2002) used a  principal component analysis  of galaxy
spectra   taken  from   the  2dFGRS   to  obtain   a   {\it  spectral}
classification scheme.  They introduced the parameter $\eta$, a linear
combination  of the two  most significant  principal components,  as a
galaxy type classification measure. As shown by Madgwick \etal (2002),
$\eta$  follows a  bimodal distribution  and can  be interpreted  as a
measure  for  the  current  star   formation  rate  in  each  galaxy.  
Furthermore $\eta$  is well  correlated with {\it  morphological} type
(Madgwick  2002).   In  what   follows  we  adopt  the  classification
suggested by Madgwick  \etal and classify galaxies with  $\eta < -1.4$
as `early-types' and  galaxies with $\eta \geq -1.4$  as `late-types'. 
Each galaxy in our MGRS is  assigned a type (early or late), using the
method described in Yang \etal  (2004a).

As  shown in  van den  Bosch \etal  (2003a), the  observed correlation
lengths of  early and late type  galaxies in the  2dFGRS indicate that
the former are preferentially hosted by massive haloes. However, these
data  alone  do  not  contain  sufficient  information  to  accurately
constrain the segregation  of the galaxy population in  early and late
types.   Here we use  the galaxy  groups selected  from the  2dFGRS to
directly  constrain the  occupation  statistics of  both populations.  
Fig.~\ref{fig:types} plots  the fraction  of early-type galaxies  as a
function of halo  mass. Results are shown separately  for all (central
plus satellite)  galaxies (left-hand  panel) and for  central galaxies
only (right-hand panel). The solid, dotted and dashed lines correspond
to the `mass-limited'  group samples V2, V3 and  V4, respectively (see
Table~1).   Among the  total  population, the  fraction of  early-type
galaxies   increases   from  about   25\%   in   haloes  with   $M\sim
10^{12}h^{-1}M_{\odot}$   to  about   80\%  in   haloes   with  $M\sim
10^{15}h^{-1}M_{\odot}$.   Among the  central  galaxy population,  the
increase  of the fraction  of early  types with  mass is  stronger: in
haloes with  $M \gta 10^{14}  h^{-1} M_{\odot}$ virtually  all central
galaxies  are early  types.  As  a comparison,  the open  circles with
errorbars in  Fig.~\ref{fig:types} show the results  obtained from the
`mass-limited'  group samples  (V2 in  Table 1)  constructed  from the
fiducial MGRSs.  The model  predictions agree reasonably well with the
observational  results,  indicating   that  our  model  for  splitting
galaxies in early-  and late-types (see van den  Bosch \etal 2003a) is
sufficiently accurate.

\begin{figure*}
\centerline{\psfig{figure=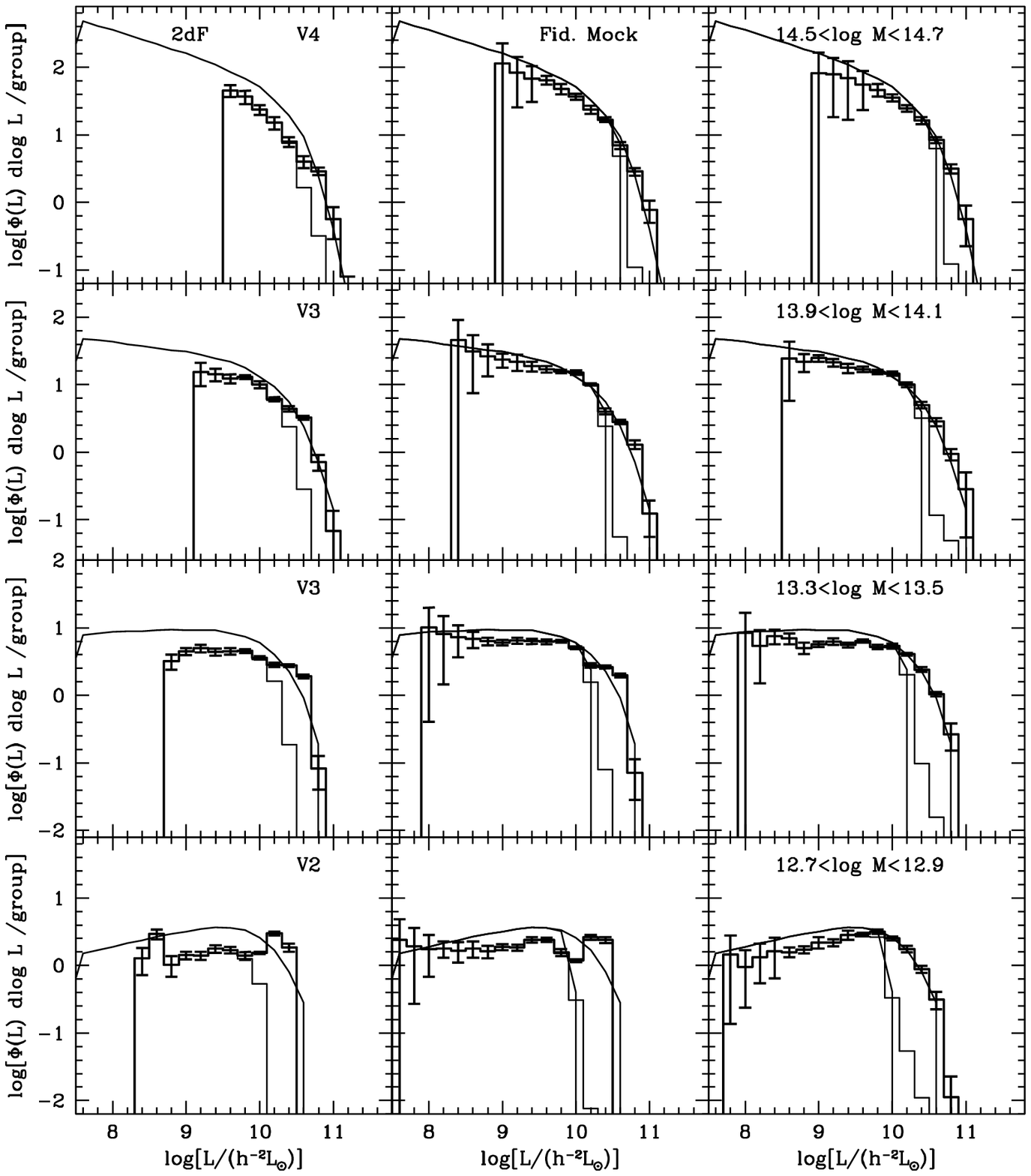,width=\hdsize}} 
\caption{The conditional luminosity functions for groups in different
  mass  bins, as  indicated.   Results are  shown  separately for  all
  (central  plus  satellite)  galaxies  (the  broader  histogram)  and
  satellite  galaxies  (the  narrower histogram),  respectively.   The
  left-hand panels are for 2dF groups, while the middle panels are for
  groups in the  fiducial MGRSs.  In all these  cases, halo masses are
  based on  the rank-ordering of the group  $L_{18}$ luminosities.  To
  test the impact of the error in the $L_{18}$-$M$ conversion, we show
  in the right  panels the CLFs obtained from  the fiducial MGRSs with
  groups binned  according to their {\it true}  halo masses. Errorbars
  in all panels  are obtained from the 1-$\sigma$  scatter among the 8
  fiducial  MGRSs.  As  a comparison,  the solid  curves  indicate the
  input  CLFs (which  are of  Schechter  form) used  to construct  the
  fiducial MGRSs.}
\label{fig:2dF_CLF}
\end{figure*}
\begin{figure*}
\centerline{\psfig{figure=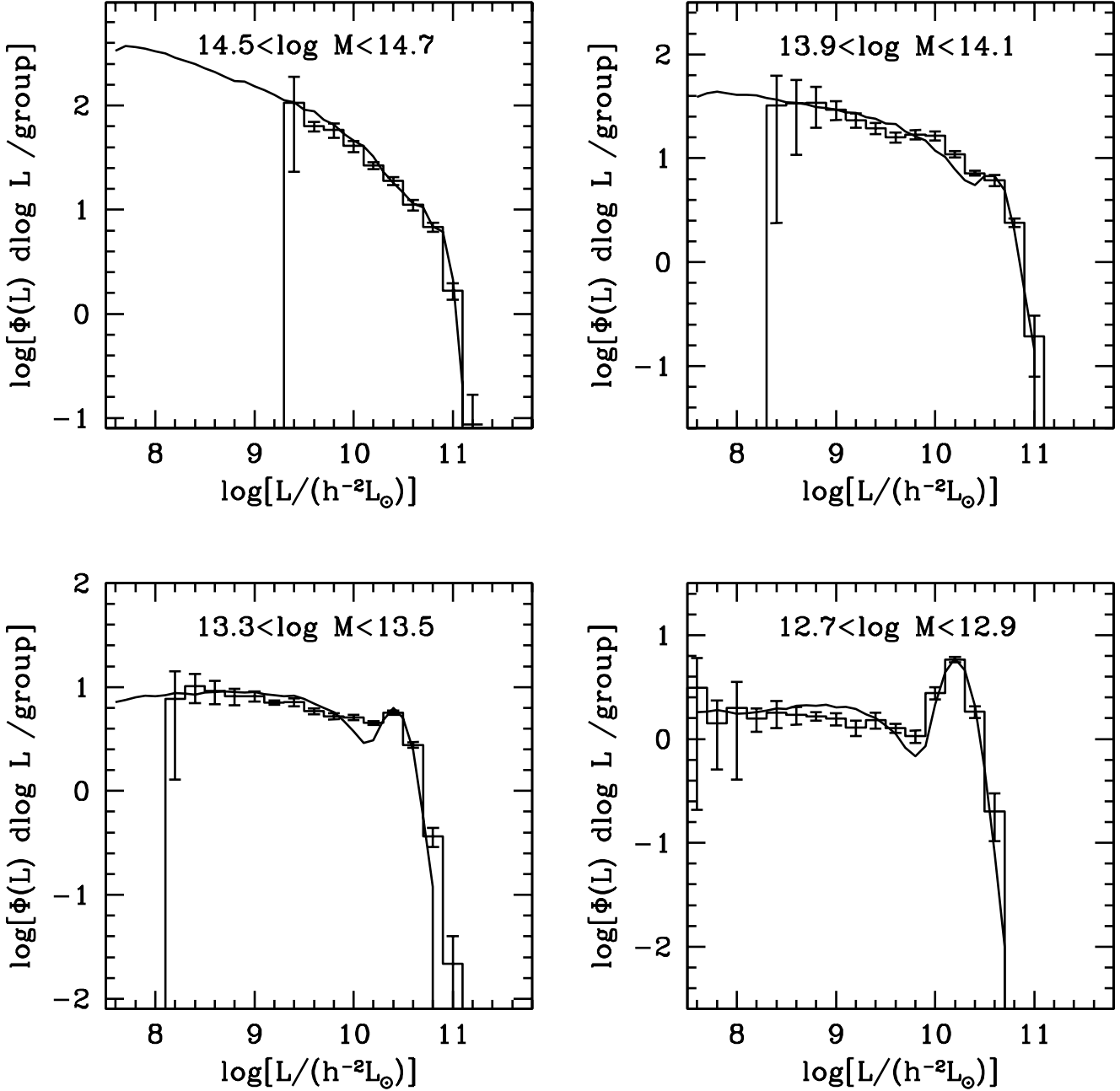,width=\hdsize}} 
\caption{
The conditional luminosity functions for groups in different
mass  bins, as  indicated. The solid curves show the input CLFs, 
which contain peaks at the bright end to mimic the conditional
baryonic mass functions found by Zheng \etal (2004).  
The histograms are the CLFs recovered from the groups selected 
from the MGRSs constructed with these input CLFs.
Errorbars in all panels are based on the 1-$\sigma$ 
scatter among 8 MGRSs.} 
\label{fig:CLF_bump}
\end{figure*}
\begin{figure*}
\centerline{\psfig{figure=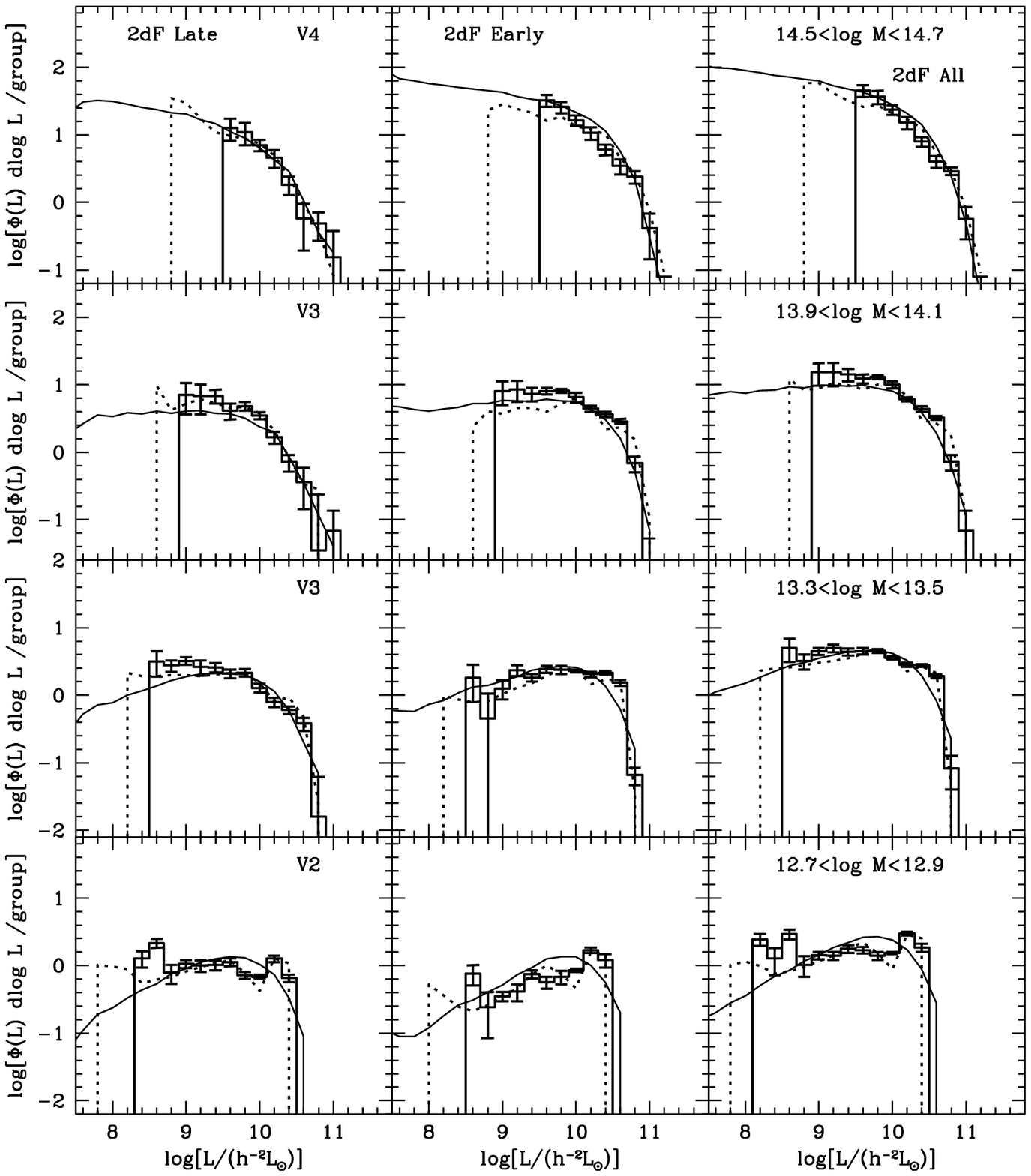,width=\hdsize}} 
\caption{The conditional luminosity functions for 2dF groups in 
  different  mass  bins (histograms), as  indicated. Results are  
  shown  separately for late-type galaxies (left panels), 
  early-type galaxies (middle panels) and all (early- and late-type)
  galaxies (right  panels). Errorbars in all panels are obtained 
  from the 1-$\sigma$ scatter among 8 MGRSs.  The solid curves  
  indicate the input CLFs used to construct the MGRSs
  with $(M/L)_{cl}=900h$. The dotted lines are the CLFs 
  recovered from the groups in these MGRSs, with halo masses
  estimated from the $L_{18}$ ranking.}
\label{fig:2dF_CLF2}
\end{figure*}

\section{The Conditional Luminosity Function}
\label{sec:CLF}

Thus  far our  discussion has  only  focused on  the occupation  {\it
  number} of galaxy groups (dark matter haloes). We now use the groups
in  the  2dFGRS  to  directly  determine  the  conditional  luminosity
function, $\Phi(L \vert M)$, which specifies the number of galaxies in
haloes {\it as a function of luminosity}. 

\subsection{Direct measurement of the CLF from 2dFGRS groups}
\label{sec:clf}

The  left-hand panels of  Fig.~\ref{fig:2dF_CLF} show  the conditional
luminosity functions (CLFs) of galaxies for 2dFGRS groups of different
masses.  For comparison, the contributions from satellite galaxies are
shown separately.  These CLFs  have been obtained directly by counting
galaxies in  groups.  For  a given galaxy  luminosity $L$, there  is a
limiting redshift, $z_L$, beyond which galaxies with such a luminosity
are not included in the sample.  In order to estimate the CLF, $\Phi(L
\vert M)$, at given $L$ and $M$  we only use groups with mass $M$ that
are within  the redshift limit $z_L$.  The  errorbars shown correspond
to 1-$\sigma$  fluctuations among 8 independent MGRSs  and reflect the
expected errors  due to cosmic  variance.  To test the  reliability of
the measurements, we compare in the middle panels the model input CLFs
(solid curves)  with those obtained  from the mock group  samples. The
CLFs recovered from  the groups in the MGRS agree  well with the model
input  down  to halo  masses  of  $M\sim  10^{13.3}\msunh$.  For  less
massive  haloes,  however, there  is  a  significant discrepancy.   In
particular, the  CLFs determined from  the groups seem to  predict too
few faint  satellite galaxies, and  too many bright central  galaxies. 
There  are two  potential  sources for  these  discrepancies: (i)  the
inaccuracy of our group finder for poor systems; (ii) the error in the
$L_{18}$ to $M$ conversion.  In order to test these possibilities, the
right-hand  panels of  Fig.~\ref{fig:2dF_CLF} plot  the CLFs  for MGRS
groups  binned according  to true  halo mass  instead of  the $L_{18}$
ranking.  This solves  the problem at the bright  end, suggesting that
this  particular  discrepancy  owes  to  errors  in  the  $L_{18}$-$M$
conversion, but still results in too few faint galaxies. This reflects
the incompleteness  of our group  finder (see Yang \etal  2004b). Note
that, in  low-mass haloes where  $L_{18}$ is dominated by  the central
galaxy, the $L_{18}$-$M$ conversion produces an artificial peak in the
CLF at the  bright end (see the bottom middle panel).   Such a peak is
also seen in  the observational data (the bottom  left panel), but our
tests show that it is doubtful that this is a real feature of the CLF.

These results are interesting in light of the recent findings by Zheng
\etal  (2004), who  computed the  conditional baryonic  mass functions
(hereafter  CMF) of  galaxies in  the semi-analytical  models  of Cole
\etal (2000).  In haloes with $12.5 \lta {\rm log}[M/\Msun] \lta 14.0$
these CMFs reveal a clear peak  due to the central galaxies. Before we
can claim that this is  inconsistent with our results presented above,
we need to show that if a  peak is present in the CLF, our analysis is
able to  recover it. To test  this, we construct eight  MGRSs based on
CLFs that contain  artificial peaks at the bright  end, similar to the
CMFs in  Zheng \etal (2004).   The input CLFs  are shown as  the solid
curves   in   Fig.~\ref{fig:CLF_bump}.    We   then  apply   all   the
observational  selections  to  these  MGRSs and  select  groups  using
exactly the  same method as that  for the real  groups.  The recovered
CLFs in different mass  ranges are shown in Fig.~\ref{fig:CLF_bump} as
the histograms  (the errorbars are 1-$\sigma$ scatter  among the eight
MGRSs).   Comparing  these with  the  model  input,  we see  that  our
analysis is able to recover prominent peaks in the CLFs, although weak
and sharp  features may be smeared  out.  Thus, if the  CLF for haloes
with  $M\la 10^{13.5}\msunh$  indeed contained  peaks as  prominent as
those  predicted by  the  semi-analytical model  mentioned above,  our
analysis would  have easily revealed them. Therefore  we conclude that
the  true CLF,  as  extracted from  the  2dFGRS, does  not reveal  any
prominent  peaks.  This  suggests  a problem  for the  semi-analytical
models of Cole \etal (2000), although we caution that any disagreement
between the  CLF (based on  luminosity in the  photometric $b_J$-band)
and  the CMF  (based  on  baryonic mass)  should  be interpreted  with
extreme care.

\subsection{Comparison with CLF Models}
\label{sec:mocks}

Having measured the CLFs from the groups in the 2dFGRS, we now turn to
a comparison  with the  results obtained from  the MGRSs and  with the
actual input CLFs used to  construct them.  Recall that the input CLFs
are based on matching the observed luminosity function of galaxies and
the  luminosity-dependence  of the  correlation  length  of galaxies.  
Comparing  the   2dF  results  shown   in  the  left-hand   panels  of
Fig.~\ref{fig:2dF_CLF}  with the  results obtained  from  the fiducial
MGRSs (the middle  panels), we see that the  observed CLFs have shapes
that  are  similar  to  our   model  predictions,  but  with  a  lower
amplitudes.   This is  simply  another reflection  of the  discrepancy
between the 2dFGRS and our fiducial CLF model regarding the abundances
of satellite galaxies (see  Section~\ref{sec:central} and also van den
Bosch  \etal  2004b).   Similar  discrepancies  have  previously  been
noticed from  the pairwise  peculiar velocity dispersions  (Yang \etal
2004a)  and the  multiplicity function  of galaxy  groups  (Yang \etal
2004b).   As  shown in  these  studies,  these discrepancies  indicate
either a relatively  high mass-to-light ratio on cluster  scales, or a
relatively low value of $\sigma_8$.  To test how the corresponding CLF
models  compare to the  CLF derived  directly from  the groups  in the
2dFGRS, the dotted lines in Fig.\,\ref{fig:2dF_CLF2} indicate the CLFs
obtained from  the MGRSs  with $(M/L)_{\rm cl}  \simeq 900 h  \MLsun$. 
Results  are  shown  separately  for early-type,  late-type,  and  all
galaxies.  Clearly,  this model is  in much better agreement  with the
2dFGRS than the  fiducial model. Note that the  model also very nicely
reproduces the CLFs  of the early- and late-type  galaxies separately. 
As for  our fiducial mocks, the  input CLFs (solid  smooth curves) are
well   reproduced  by  the   MGRSs  for   haloes  more   massive  than
$10^{13.3}\msunh$, while  for less massive haloes a  small peak arises
and the groups become incomplete at the faint end.

Unfortunately, we  do not  have a set  of numerical simulations  for a
$\Lambda$CDM  cosmology  with  $\sigma_8=0.7$,  so  that  we  can  not
construct  corresponding  MGRSs  (but  see  Yang \etal  2004a  for  an
approximate  method).   However,   numerous  tests  discussed  in  our
previous work suggests that  this model is virtually indistinguishable
from  that  with  $\sigma_8=0.9$  and  $(M/L)_{\rm cl}  \simeq  900  h
\MLsun$. We therefore conclude  that the CLFs determined directly from
the groups in 2dFGRS provide further support for both models as viable
descriptions of the galaxy-dark matter connection.

\section{Conclusions}
\label{sec:conclusion}

Using the galaxy  group catalogue constructed from the  2dFGRS by Yang
\etal (2004b), we have investigated various aspects regarding the halo
occupation statistics  of galaxies in  the 2dFGRS.  This is  the first
time  the  halo occupation  distribution  in  real  galaxy systems  is
examined in such  detail, and has resulted in  a number of interesting
results that  shed light on  the connection between galaxies  and dark
matter haloes.

In order to estimate halo masses associated with the galaxy groups, we
have  ranked groups according  to their  group luminosity.   Under the
ansatz that the group luminosity is tightly correlated with halo mass,
this is similar  to mass ranking, and one can  use the mean separation
between   the  groups  above   a  given   ranking  to   determine  the
corresponding  halo mass.   Since  any stochasticity  in the  relation
between group  luminosity and halo  mass causes errors in  the derived
group  masses, it  is essential  to use  mock galaxy  redshift surveys
(MGRSs)  to properly  interpret the  results.  In  this study  we used
MGRSs  constructed  using the  conditional  luminosity function  (CLF)
which has been constrained  by the abundance and clustering properties
of the galaxies in the 2dFGRS.

The first statistic we have investigated is the mean occupation number
of galaxies  above a given luminosity  limit.  Using the  MGRS we have
shown that  this statistic  can be determined  from the  galaxy groups
extremely  reliably,  except for  low  mass  haloes  where $\langle  N
\rangle  \simeq   1$.   Here  the  stochasticity   in  the  occupation
statistics causes a systematic error that mimics a flat shoulder and a
sharp break in the derived $\langle N \rangle_M$.  Yet, the comparison
between  2dFGRS  and  MGRS,  both   of  which  suffer  from  the  same
systematic,  is  meaningful  and   allows  one  to  test  whether  the
occupation statistics  used to construct  the MGRS (i.e., the  CLF and
its sampling  strategy) are in agreement  with the data.   In terms of
the  mean   occupation  numbers,  we  find  that   our  fiducial  MGRS
overestimates $\langle N \rangle$ in  high mass haloes with respect to
the  2dFGRS.   This overabundance  of  satellite  galaxies in  massive
haloes was previously noticed in  van den Bosch \etal (2004b) and Yang
\etal  (2004b), and  indicates that  either clusters  have  an average
mass-to-light ratio of $(M/L)_{\rm  cl} \simeq 900 h \MLsun$ (compared
to $(M/L)_{\rm cl} = 500 h \MLsun$ in our fiducial model), or that the
power-spectrum normalization is  relatively low; $\sigma_8 \simeq 0.7$
rather  than $0.9$.  Similar  conclusions were  reached by  Yang \etal
(2004a)  from a  detailed analysis  of the  pairwise-peculiar velocity
dispersions of galaxies in the 2dFGRS.

In addition to the mean, we  have also investigated the scatter in the
occupation  statistics, using  the ratio  $\sigma(N)  /\sqrt{\langle N
  \rangle}$.  In massive  haloes we  find this  ratio to  be  close to
unity, indicating  that the occupation distribution $P(N  \vert M)$ is
close  to  Poissonian.  In  low  mass,  haloes,  however, there  is  a
pronounced minimum  indicating a $P(N \vert M)$  that is significantly
narrower  than  a Poisson  distribution.   We  have  shown that  these
findings are in excellent agreement  with our fiducial MGRSs, but only
if we sample the luminosity of  the brightest galaxy in each halo in a
somewhat  deterministic way.   Without such  special  treatment, i.e.,
when  drawing all  luminosities  at  random form  the  CLF, the  ratio
$\sigma(N)/\sqrt{\langle N  \rangle}$ is  no longer in  agreement with
that of  the 2dFGRS.  These  findings suggest that (i)  the occupation
statistics  of central  galaxies are  sub-Poissonian,  indicating some
deterministic  behavior in galaxy  formation, and  (ii) the  number of
satellite  galaxies  above  a   certain  luminosity  limit  follows  a
Poissonian  distribution.   This  is  in excellent  agreement  with  a
scenario in  which satellite galaxies are associated  with dark matter
subhaloes,  which, as  shown  by Kravtsov  \etal  (2004), also  follow
Poissonian occupation statistics.

The mean luminosity of the  central galaxies, $L_c$, is found to scale
with halo  mass as  $L_c\propto M^{2/3}$ for  haloes with masses  $M <
10^{13}h^{-1}\msun$,  and  as $L_c\propto  M^{1/4}$  for more  massive
haloes.   At the  low-mass end,  this is  in excellent  agreement with
results  based on  galaxy-galaxy  weak lensing,  which  imply that  $M
\propto L_c^{1.5}$  (e.g.  Yang  \etal 2003a; Guzik  \& Seljak  2002). 
The  characteristic  break   at  $M  \simeq  10^{13}  h^{-1}M_{\odot}$
indicates the existence of a characteristic scale in galaxy formation,
thought  to  be  associated  with  the transition  from  effective  to
ineffective cooling (e.g., White  \& Rees 1978; Dekel 2004).  Although
not  directly revealed  by our  galaxy groups,  another characteristic
mass,  $M\simeq 10^{11}h^{-1}\msun$,  can  be inferred  from our  CLFs
obtained from the 2dFGRS.   Below this mass, star formation efficiency
decreases  rapidly  with  decreasing  halo  mass,  presumably  due  to
feedback  from   supernovae.  We  have  also   investigated  the  full
distribution  of central  luminosity; $P(L_c  \vert M)$.   Although it
appears  log-normal, detailed  tests  show that  the group  luminosity
ranking  used to  estimate  halo masses  causes  systematic errors  in
$P(L_c \vert  M)$ (though the  mean is not affected).   The comparison
with  the  MGRS, however,  is  still  meaningful  and shows  excellent
agreement,  providing further  support for  the CLF  and  its sampling
strategy.

In addition to a split in central and satellite galaxies, we have also
divided the  population in early- and late-type  galaxies. The central
galaxies in  low-mass haloes are  found to be predominantly  late type
galaxies,  while those  in massive  haloes are  almost  entirely early
types.  This  is in  good  agreement  with  the occupation  statistics
obtained from an  analysis of the clustering properties  of early- and
late-type galaxies (van den Bosch \etal 2003a).

Using the 2dF groups, we have also measured the conditional luminosity
function  directly.  Although the  CLF of  central galaxies  is fairly
narrow, the presence of central galaxies  does not show up as a strong
peak at the bright end of the total CLF. In fact, over the entire halo
mass range  that can  be reliably probed  with the present  data (from
$\sim 10^{13.3}h^{-1}\msun$  to $\sim 10^{14.7}h^{-1}\msun$),  the CLF
is well  fit by  a Schechter function.   This supports  the assumption
regarding  the  shape  of the  CLF  made  in  our previous  work,  but
disagrees  with  the  conditional  baryonic  mass  function  (CMF)  in
semi-analytical models  of galaxy formation.  As shown  by Zheng \etal
(2004),  the latter  reveals  a  pronounced peak  due  to the  central
galaxies.  We caution, however,  that any disagreement between the CLF
(based on luminosity in the photometric $b_J$-band) and the CMF (based
on baryonic mass) should be interpreted with care.

The CLFs  obtained from the  galaxy groups in  the 2dFGRS are  in good
agreement with the CLF model based on matching the observed luminosity
function  and large-scale  clustering  properties of  galaxies in  the
$\Lambda$CDM  concordance  cosmology.  It  indicates  that this  model
provides an  accurate description  of the connection  between galaxies
and  dark matter  haloes,  with the  condition  that either  $\sigma_8
\simeq 0.7$, or, if $\sigma_8=0.9$, that clusters have a mass-to-light
ratio that  is significantly higher than typically  found.  Finally we
point  out  that,  with  the  completion of  the  SDSS,  the  analysis
presented here can be naturally extended to include a wider variety of
intrinsic properties of individual galaxies (in addition to luminosity
and  type), to investigate  the occupation  statistics as  function of
color, surface  brightness, AGN activity, etc.  The  results from such
analyzes  will provide  unprecedented constraints  on how  galaxies of
different physical properties form  in dark matter haloes of different
masses.

%%%%%%%%%%%%%%%
% Acknowledgements
%%%%%%%%%%%%%%%

\section*{Acknowledgement}

Numerical  simulations used  in this  paper  were carried  out at  the
Astronomical Data Analysis Center  (ADAC) of the National Astronomical
Observatory,  Japan. We  thank  the  2dF team  for  making their  data
publicly  available  and  the  anonymous  referee  for  many  valuable
comments.  FB thanks Peder Norberg  for vivid discussions on groups in
the 2dFGRS.

%%%%%%%%%%%%%%%
% Bibliography
%%%%%%%%%%%%%%%

\label{lastpage}

\end{document}

%% file: macro1.tex
\newcommand{\etal}{{et al.~}}
\newcommand{\lta}{\la}
\newcommand{\gta}{\ga}

\newcommand{\kmsmpc}{\>{\rm km}\,{\rm s}^{-1}\,{\rm Mpc}^{-1}}

\newcommand{\Msun}{\>{\rm M_{\odot}}}
\newcommand{\msun}{\>{\rm M_{\odot}}}
\newcommand{\Lsun}{\>{\rm L_{\odot}}}
\newcommand{\MLsun}{\>({\rm M}/{\rm L})_{\odot}}

\newcommand{\beq}{\begin{equation}}
\newcommand{\eeq}{\end{equation}}

\newcommand{\mpch}{\>h^{-1}{\rm {Mpc}}}

\newcommand{\msunh}{\>h^{-1}\rm M_\odot}

\newcommand{\apj}{ApJ}

\newcommand{\mnras}{MNRAS}

\newdimen\hssize
\newdimen\hdsize 